\documentclass[onecolumn,fleqn,usenatbib]{mnras}

\usepackage{newtxtext,newtxmath}
\usepackage[T1]{fontenc}

\usepackage{amsmath}
\usepackage{bm}

\numberwithin{equation}{section}

\usepackage{graphicx}

\usepackage[capitalise,compress]{cleveref}
\newcommand{\crefrangeconjunction}{--}
\newcommand{\creflastconjunction}{, and\nobreakspace}
\crefname{section}{Sec.}{Secs.}
\Crefname{section}{Section}{Sections}

\DeclareRobustCommand{\VAN}[3]{#2}
\let\VANthebibliography\thebibliography
\def\thebibliography{\DeclareRobustCommand{\VAN}[3]{##3}\VANthebibliography}

\usepackage{soul}

\newcommand{\dd}{\mbox{d}}

\title[The dynamical tides of spinning Newtonian stars]{The dynamical tides of spinning Newtonian stars}

\author[P. Pnigouras et al.]{%
    P. Pnigouras,$^{1,2,3}$\thanks{pantelis.pnigouras@ua.es}
    F. Gittins,$^{4}$
    A. Nanda,$^{5,6}$
    N. Andersson$^{4}$
    and D. I. Jones$^{4}$
    \\
    $^{1}$Departamento de F\'isica Aplicada, Universidad de Alicante, Campus de San Vicente del Raspeig, Alicante 03690, Spain\\
    $^{2}$Dipartimento di Fisica, ``Sapienza'' Universit{\`a} di Roma \& Sezione INFN Roma1, Piazzale Aldo Moro 2, Roma 00185, Italy\\
    $^{3}$Department of Physics, Aristotle University of Thessaloniki, Thessaloniki 54124, Greece\\
    $^{4}$Mathematical Sciences and STAG Research Centre, University of Southampton, Southampton SO17 1BJ, United Kingdom\\
    $^{5}$Indian Institute of Science Education and Research, Dr. Homi Bhabha Road, Pashan, Pune 411 008, India\\
    $^{6}$Research Center for the Early Universe (RESCEU), University of Tokyo, Tokyo 113-0033, Japan
}
             
\date{Accepted XXX. Received YYY; in original form ZZZ}

\pubyear{2023}

\begin{document}
\label{firstpage}
\pagerange{\pageref{firstpage}--\pageref{lastpage}}
\maketitle


\begin{abstract}
    We carefully develop the framework required to model the dynamical tidal response of a spinning neutron star in an inspiralling binary system, in the context of Newtonian gravity, making sure to include all relevant details and connections to the existing literature. The tidal perturbation is decomposed in terms of the normal oscillation modes, used to derive an expression for the effective Love number which is valid for any rotation rate. In contrast to previous work on the problem, our analysis highlights subtle issues relating to the orthogonality condition required for the mode-sum representation of the dynamical tide and shows how the prograde and retrograde modes combine to provide the overall tidal response. Utilising a slow-rotation expansion, we show that the dynamical tide (the effective Love number) is corrected at first order in rotation, whereas in the case of the static tide (the static Love number) the rotational corrections do not enter until second order.
\end{abstract}




\section{Introduction} \label{sec:Introduction}

The emergence of gravitational-wave astronomy, beginning with the historic observation of the coalescing black-hole binary GW150914 \citep{2016PhRvL.116f1102A}, has brought with it a renewed interest in compact-object binaries. Since the inaugural gravitational-wave detection, ground-based instruments have captured a variety of compact-binary mergers, the overwhelming majority comprising binary black holes  \citep{2019PhRvX...9c1040A, 2021PhRvX..11b1053A, 2021arXiv211103606T}, but also including two binary neutron stars \citep{2017PhRvL.119p1101A, 2020ApJ...892L...3A} and two neutron star-black hole binaries \citep{2021ApJ...915L...5A}.


The gravitational-wave signal from an inspiralling binary hosting at least one neutron star differs slightly from that of two black holes. By virtue of being extended material bodies, neutron stars are tidally deformed when immersed in an external gravitational field. As the binary coalesces, the deformability of the stellar fluid imprints finite-size corrections on the signal, which manifest as a dephasing in the waveform compared to an equivalent-mass black-hole binary \citep{1992ApJ...398..234K, 1992ApJ...400..175B}. These finite-size effects are sensitive to the internal structure of the material body and are usually parametrised through the so-called \textit{tidal Love numbers} \citep{2008ApJ...677.1216H}. Thus, a particularly tantalising prospect for gravitational-wave observations of neutron stars is the opportunity to obtain constraints on the elusive dense nuclear-matter equation of state \citep{2008PhRvD..77b1502F, 2010PhRvD..81l3016H, 2013PhRvD..88j4040M, 2013PhRvL.111g1101D, 2020PhRvC.102e5801K, 2021Symm...13..183K, foundations1020017}. Indeed, such a measurement was attempted for the celebrated multimessenger event GW170817, which---through a lack of evidence for tidal deformations in the signal---delivered an astrophysically interesting upper bound, favouring softer equations of state, \textit{i.e.}, models that produce smaller, less deformable stars (\citealp{2017PhRvL.119p1101A, 2017ApJ...850L..34B, 2018PhRvL.121i1102D, 2018PhRvL.121p1101A, 2018PhRvL.120z1103M, 2018PhRvL.120q2703A, 2018ApJ...857L..23R, 2018ApJ...852L..29R, 2018PhRvC..98c5804M, 2018PhRvC..98d5804T, 2019PhRvL.123n1101F, 2019EPJA...55...50R}; see also reviews by \citealt{2019JPhG...46l3002G} and \citealt{2020GReGr..52..109C}). Such constraints are expected to improve significantly with next-generation terrestrial detectors, like the Einstein Telescope and the Cosmic Explorer \citep{2022PhRvL.128j1101P}, that are currently under development.

According to current (electromagnetic) binary pulsar observations, the neutron-star mass distribution peaks at about $1.4\,M_\odot$ (with $M_\odot$ being the solar mass; \citealt{2013ApJ...778...66K}), whereas the most massive neutron star observed to date (known as PSR J0740+6620) has a mass of $2.08\pm 0.07\,M_\odot$ \citep{2020NatAs...4...72C,2021ApJ...915L..12F}, already excluding a number of theoretically proposed (soft) equations of state. Constraints on the neutron-star radius (and thus on the equation of state) from electromagnetic observations have significantly improved with the NASA NICER mission (``Neutron Star Interior Composition Explorer"), which recently provided radius estimates for two pulsars (including PSR J0740+6620), hinting at a preference for stiffer equations of state, compared to the measurements obtained from GW170817 \citep{2019ApJ...887L..21R, 2019ApJ...887L..24M, 2020ApJ...893L..21R, 2021ApJ...918L..27R, 2021ApJ...918L..28M}. In addition, terrestrial experiments, such as PREX-II that measured the neutron skin thickness of lead ($^{208}$Pb), may also be used to constrain the neutron-star radius \citep{2021PhRvL.126q2503R}, thus complementing astronomical observations.

The \textit{static} tidal deformations of spherically symmetric (non-rotating) fluid bodies are well understood in both Newtonian \citep{2014ARA&A..52..171O, 2014grav.book.....P} and relativistic gravity \citep{2008ApJ...677.1216H, 2009PhRvD..80h4018B, 2009PhRvD..80h4035D, 2015PhRvD..91j4026L, 2018PhRvD..98l4023P}, and it has become routine to compute high-precision, perfect-fluid stellar models using realistic equations of state \citep{2010PhRvD..81l3016H, 2010PhRvD..82b4016P}. Furthermore, there has been work on introducing more complexity into the neutron-star models, such as elastic crusts \citep{2020PhRvD.101j3025G} and superfluidity \citep{2018PhRvD..98h4010C, 2020PhRvD.101f4016D, 2021Univ....7..111Y}. In addition, the components of binary systems are expected to be rotating (albeit perhaps slowly, since the stars involved are likely to be old).%
\footnote{Extracting the angular momentum of binary components from a gravitational waveform is a challenging task \citep{1995PhRvD..52..848P, 2013PhRvD..87b4035B, 2014PhRvL.112y1101V, 2015ApJ...798L..17C, 2016ApJ...825..116F, 2016PhRvX...6d1015A, 2017PhRvD..95f4053V, 2020PhRvR...2d3096P} due  to the fact that, during inspiral, the leading-order post-Newtonian influence on the evolution of the system arises as a mass-weighted combination of the two spins \citep{2001PhRvD..64l4013D, 2014LRR....17....2B, 2016PhRvD..93h4042P, 2018PhRvD..98h3007N, 2020ApJ...899L..17Z}.}
For this reason, there has been a substantial effort in studying the relativistic deformations of spinning compact objects in the presence of static tidal fields, initially focusing on Kerr black holes (\citealp{2006PhRvD..73b4010Y, 2014PhRvD..90l4039O, 2015PhRvD..91d4004P, 2021PhRvD.104b4013C}; showing that their multipolar structure is unaffected by the presence of a static tidal field, extending the corresponding result for Schwarzschild black holes; \citealp{2009PhRvD..80h4018B, 2009PhRvD..80h4035D, 2015PhRvL.114o1102G, 2021PhRvD.104j4062P}) and more recently considering the material case \citep{2015PhRvD..91j4018L, 2015PhRvD..92b4010P, 2015PhRvD..92l4003P, 2015PhRvD..92l4041L, 2017PhRvD..95l4058L, 2021PhRvD.104d4052C, 2022PhRvD.106b4011C}. Currently, all such calculations are at the level of slowly rotating fluid bodies.

Treating the exterior field as static is an appropriate approximation for binaries in the \textit{adiabatic regime}, where the two components are widely separated and the evolution is slow. However, as the bodies get closer to one another, the dynamics become important \citep{2008PhRvD..77b1502F, 2012PhRvD..86d4032M}. Our understanding of \textit{dynamical tides}  is comparatively less well developed. The phenomenology of the problem is well-established---the standard strategy in Newtonian gravity is to represent the tidal response as a sum over the star's oscillation modes---but progress towards including more realistic physics has been somewhat limited. When it comes to rotational effects, the notable recent progress on modelling tides in gas planets \citep{2021PSJ.....2...69I, 2021PSJ.....2..122L, 2022ApJ...925..124D, 2022PSJ.....3...11I, 2022PSJ.....3...89I} is due to the \textit{Juno} and \textit{Cassini} missions observing Jupiter and Saturn, respectively. Hosting several moons and rotating relatively rapidly, the large gas planets are suitable environments for the study of dynamical tidal effects. In fact, the \textit{Juno} spacecraft has reported a tidal Love number for Jupiter $4\%$ smaller than the hydrostatic value \citep{2020GeoRL..4786572D}, a discrepancy which can in principle be resolved by considering dynamical effects in the tidal response \citep{2021PSJ.....2...69I, 2021PSJ.....2..122L, 2022PSJ.....3...89I}. Meanwhile, the impact of relativity, essential for binary neutron-star inspirals and relevant for gravitational-wave astronomy, is tricky to establish. Basically, the relativistic mode-sum approach has not yet been developed to the level required.%
\footnote{There are compelling reasons to expect the set of quasinormal modes to be incomplete (due to the presence of gravitational waves) and this complicates the tidal problem, at least in principle.}
Nevertheless, there has been recent work on incorporating dynamical tides into the effective-one-body framework \citep{2016PhRvL.116r1101H, 2016PhRvD..94j4028S}, as well as in gravitational waveform models for both non-rotating \citep{2019PhRvD.100b1501S} and (slowly) rotating neutron stars \citep{2021PhRvR...3c3129S}. In addition, the prospect of performing neutron-star seismology studies using dynamical tides has been explored \citep{2020NatCo..11.2553P}---highlighting the importance of the problem for next-generation gravitational-wave instruments, as neglecting dynamical tides might lead to significant errors in neutron-star parameter estimation \citep{2022PhRvL.129h1102P, 2022PhRvD.105l3032W, 2022PhRvD.106f4052K}. However, useful progress can still be made in Newtonian gravity \citep{1994MNRAS.270..611L, 1994ApJ...426..688R, 1995MNRAS.275..301K, 1995ApJ...449..294K, 2017MNRAS.464.2622Y, 2017MNRAS.470..350Y, 2020PhRvD.101h3001A, 2021MNRAS.503..533A, 2021MNRAS.504.1273P, 2022MNRAS.514.1494P}, where the effects of rotation have been studied in the context of resonant mode excitation by the orbital motion \citep{1997ApJ...490..847L, 1999MNRAS.308..153H, 2006PhRvD..74b4007L, 2017PhRvD..96h3005X}, and post-Newtonian theory \citep{2007PhRvD..75d4001F, 2017PhRvD..95d4023P, 2019PhRvD.100f3001V, 2020PhRvD.101j4028P, 2020PhRvD.102f4059P, 2020PhRvD.101f4003B, 2021PhRvD.103f4023P}.

In previous work~\citep{2020PhRvD.101h3001A} we developed a formalism which describes the tidal perturbation in terms of the stellar oscillation modes and provides the contribution of each mode to the \textit{effective} Love number, which includes dynamical effects. Such a description is helpful in order to gain insight into the physics of the stellar interior, given that the properties of the various classes of modes strongly depend on the internal structure of the star (the basic principle of \textit{asteroseismology}). Neutron stars possess a rich spectrum of normal oscillation modes (\textit{e.g.}, see \citealt{1988ApJ...325..725M, 1991ApJ...372..573S, 1996PhRvL..77.4134A, 2001PhRvD..65b4001S, 2003MNRAS.338..389M, 2004PhRvD..70h4009G, 2004PhRvD..70l4015B, 2008PhRvD..78f4063G, 2012MNRAS.419..638P, 2013PhRvD..88d4052D, 2014PhRvD..90b4010G, 2015PhRvD..92f3009K, 2018MNRAS.478..167S, 2018Springer..PP, 2019gwa..book.....A, 2020PhRvL.125k1106K}). During inspiral, the tidal force will sweep through a range of frequencies until coalescence, which will result in the resonant excitation of different modes when the tidal frequency matches a mode's eigenfrequency. This is interesting, since the rapid growth of a mode during resonance may extract energy from the orbit, thus affecting the phasing of the gravitational waveform. Should these dynamical tidal effects be distinguishable, one could then perform neutron star seismology studies using gravitational waves from neutron star binaries. The implications of this have been explored previously, but the interest has been inevitably renewed in the last few years, mostly focusing on the effects that the fundamental mode ($f$ mode) has on the signal (see references on dynamical tides above). This is because the $f$ mode, being a large-scale perturbation, is expected to contribute the most to the dynamical tidal response of the star. Furthermore, its high frequency (likely above 1~kHz) implies that its resonant excitation will be relevant only late in the inspiral (with the actual resonance likely not happening before the stars merge). The merger dynamics, and the impact of the dynamical tide, are expected to remain largely unresolved until next-generation detectors become available.

In the current work, we extend the formalism developed in \citet{2020PhRvD.101h3001A} to include rotation. Generally, neutron stars close to merger are expected to be old, implying that they will have had enough time to spin down, probably rendering rotation less important dynamically. However, as we will see, the neutron star spin introduces significant changes both to the oscillation modes, on which our description of the tide is based, and to the required formalism, thus motivating this study. Using perturbation theory,  within the context of Newtonian gravity, we carefully develop the framework required to model the dynamical tide of a spinning neutron star. In particular, our analysis highlights subtle issues like the orthogonality condition required for the mode-sum representation and how the prograde and retrograde modes combine to provide the overall tidal response of a rotating star. The way we deal with these aspects is not new, but the tidal argument has (arguably) not previously been analysed at this level of coherence and detail. In fact, we have gone out of our way to make sure all relevant details and connections to the existing literature are included.

The layout of the paper is as follows: We derive the equation of motion for the tidal perturbation in \cref{sec:Tidal deformations} and discuss the properties of normal modes in \cref{sec:Normal modes of oscillation}, particularly focusing on the completeness and the orthogonality of the modes of a rotating star. The analysis draws heavily on the Lagrangian perturbation formalism developed by \citet{1978ApJ...221..937F,1978ApJ...222..281F}, with important extensions provided by \citet{2001PhRvD..65b4001S}. For completeness, we include the key arguments from these classic papers. In fact, we have made an effort to make the analysis as logical and self-contained as possible, as it should soon become clear that we need to pay attention to the fine-print detail in order to explain the tidal response of a rotating star. In \cref{sec:Mode-sum representation} we derive an expression for the effective Love number using a decomposition of the tidal perturbation in terms of the oscillation modes, valid for arbitrary rotation rates [\cref{eq:RotatingLove}].%
\footnote{The arguments leading to the expression for the Love number proceed along similar lines to \citet{2021PSJ.....2..122L} and \citet{2022ApJ...925..124D}, where the influence of rotation is studied in the dynamical tides of planets. However, as explained later on, the rigorous analysis presented here is required in order to understand precisely how individual oscillation modes contribute to the tidal response, as well as to highlight important steps and issues which previously were either missed or not adequately explained.}
We show that, contrary to the non-rotating case, one has to use a phase-space decomposition (as in \citealt{2001PhRvD..65b4001S}), simultaneously expanding both the tidal displacement and its derivative, in order to obtain uncoupled equations of motion for the modes. We also re-visit the effective Love number derivation for the case of non-rotating stars, following slightly different arguments and arriving at an expression [\cref{eq:NonRotatingLove}] equivalent to that from \citet{2020PhRvD.101h3001A}. Finally, in \cref{sec:The slow-rotation approximation} we provide a thorough description of the oscillation modes in the slow-rotation approximation, using two equivalent approaches and highlighting subtleties associated with the symmetry of the modes and their normalisation, leading up to an expression for the effective Love number, which is valid at first order in the rotation [\cref{eq:LoveNumberFirstOrderWithResonance,eq:LoveNumberFirstOrder}], and to the demonstration that the static Love number is only corrected at second order in the rotation [\cref{eq:StaticLoveNumberFirstOrder}].

We work in a coordinate basis, using Latin scripts $i, j, k, \ldots$ to denote the spatial components of a vector and adopt the Einstein summation convention, where repeated indices represent a summation. We reserve the indices $l$ and $m$ for the multipole degree and azimuthal order, respectively, of a spherical harmonic $Y_l^m$. Greek indices $\alpha, \beta, \gamma, \ldots$ will be used to label the different oscillation modes.


\section{Tidal deformations} \label{sec:Tidal deformations}

We consider a binary system comprising a star of mass $M$ and radius $R$, uniformly rotating with angular velocity $\Omega^i$, and a companion of mass $M'$, treated as a point mass. In a spherical coordinate system $(r, \theta, \phi)$ centred on the primary $M$, with its $z$ axis aligned with the orbital angular momentum, the companion has coordinates $[D(t), \pi / 2, \psi(t)]$, where $D$ is the orbital separation between the two stars and $\psi$ is the orbital phase. The presence of the massive companion perturbs the equilibrium shape of the primary star. This is due to the gravitational potential sourced by $M'$ \citep{1977ApJ...213..183P, 1999MNRAS.308..153H},
\begin{equation}
    \chi(t, x^i) 
        = - \frac{G M'}{|x^i - D^i(t)|} 
        = - G M' \sum_{l = 2}^\infty \sum_{m = - l}^l 
            \frac{W_{l m} r^l}{D(t)^{l + 1}} Y_l^m(\theta, \phi) 
            e^{- i m \psi(t)},
    \label{eq:TidalField}
\end{equation}
where $G$ is the gravitational constant, $x^i$ is the position relative to the centre of the primary and $D^i$ is the position of the companion. It is important to note that, as the tidal potential is a real-valued function, it follows from \cref{eq:TidalField} that we must have $W_{l, - m} = (- 1)^m W_{l m}$. Moreover, due to the symmetry of the tidal potential, we have  $W_{l m} = 0$ for odd $l + m$ and otherwise
\begin{equation}
    W_{l m} 
        = \frac{4 \pi}{2 l + 1} Y_l^{m *}(\pi / 2, 0) 
        = (- 1)^{(l + m) / 2} \sqrt{\frac{4 \pi}{2 l + 1} (l - m)! (l + m)!} 
            \left[ 2^l \left( \frac{l + m}{2} \right)! 
            \left( \frac{l - m}{2} \right)! \right]^{- 1},
    \label{eq:Wlm}
\end{equation}
with the asterisk denoting complex conjugation. For a reasonable combination of binary separation and mass ratio, there exists a region of space around the primary where its gravitational potential $\Phi$ is much stronger than that of the secondary, $\chi \ll \Phi$. In this region, it is appropriate to formulate the problem in the context of perturbation theory.

We treat the tidally deformed star as a perfect fluid body and linearise the fluid equations with respect to the tidal perturbation. Then, the perturbed Euler equation, expressed in a reference frame co-rotating with the primary, is \citep{1978ApJ...221..937F, 2001PhRvD..65b4001S}
\begin{equation}
    \mathcal{E}_i 
        \equiv \partial_t^2 \xi_i + B_i^{\hphantom{i} j} \partial_t \xi_j 
            + C_i^{\hphantom{i} j} \xi_j 
        = - \nabla_i \chi,
    \label{eq:PerturbedEuler}
\end{equation}
with the operators
\begin{equation}
    B_i^{\hphantom{i} j} \xi_j = 2 \epsilon_{i j k} \Omega^j \xi^k
    \label{eq:OperatorB}
\end{equation}
and 
\begin{equation}
    C_i^{\hphantom{i} j} \xi_j 
        = \frac{\nabla_i \delta p}{\rho} 
            - \frac{\nabla_i p}{\rho^2} \delta \rho + \nabla_i \delta \Phi,
    \label{eq:OperatorC}
\end{equation}
where $\xi^i$ is the Lagrangian displacement vector, $\rho$ and $p$ are the mass density and isotropic pressure of the equilibrium fluid, respectively, $\delta$ denotes the Eulerian change of a physical quantity and $\epsilon_{i j k}$ is the antisymmetric Levi-Civita tensor associated with the Euclidean three-metric $g_{i j}$. Note that, in the non-rotating limit, $\Omega = |\Omega^i| = 0$, the operator $B_i^{\hphantom{i} j}$ vanishes and \cref{eq:PerturbedEuler} describes the tidal perturbations of a spherical fluid configuration in an inertial frame.

The Lagrangian displacement $\xi^i$ describes how fluid elements move due to the perturbation. All perturbed quantities can be related to $\xi^i$ through the rest of the fluid equations, namely, the perturbed continuity equation
\begin{equation}
    \delta \rho = - \nabla_i (\rho \xi^i),
    \label{eq:PerturbedContinuity}
\end{equation}
the perturbed Poisson's equation
\begin{equation}
    \nabla^2 \delta \Phi 
        = 4 \pi G \delta \rho 
        = - 4 \pi G \nabla_i (\rho \xi^i),
    \label{eq:PerturbedPoissons}
\end{equation}
as well as the equation of state for the perturbations
\begin{equation}
    \frac{\Delta p}{p} = \Gamma_1 \frac{\Delta \rho}{\rho},
    \label{eq:PerturbedEoS}
\end{equation}
which is expressed in terms of Lagrangian variations, related to Eulerian ones (for scalar quantities) by $\Delta p = \delta p + \xi^i \nabla_i p$. For a cold neutron star core (ignoring exotic components like hyperons and/or deconfined quarks), the adiabatic exponent $\Gamma_1$ is defined as
\begin{equation}
    \Gamma_1=\left(\frac{\partial\ln p}{\partial\ln\rho}\right)_{x_\mathrm{p}},
    \label{eq:AdiabaticExponent}
\end{equation}
with $x_\mathrm{p}$ being the proton fraction (\textit{i.e.}, the proton number density over the baryon number density). In \cref{eq:PerturbedEoS} it has been assumed that $\Delta x_\mathrm{p}=0$, namely that the composition of a perturbed fluid element remains ``frozen" during an orbital period, due to the long time scales on which the relevant $\beta$ reactions (the Urca processes) can establish chemical equilibrium \citep{1992ApJ...395..240R, 2019MNRAS.489.4043A}. \Cref{eq:PerturbedEoS} can also be written as
\begin{equation}
    \frac{\delta\rho}{\rho}=\frac{1}{\Gamma_1}\frac{\delta p}{p}-A_i\xi^i,
    \label{eq:PerturbedEoS2}
\end{equation}
where the Schwarzschild discriminant $A_i$, defined as
\begin{equation}
    A_i=\frac{\nabla_i\rho}{\rho}-\frac{1}{\Gamma_1}\frac{\nabla_i p}{p},
    \label{eq:SchwarzschildDiscriminant}
\end{equation}
determines the convective stability of the fluid. In a setup where the displaced fluid either matches or rapidly adjusts to the chemical composition of its surroundings, the Schwarzschild discriminant vanishes. Otherwise, buoyancy acts as a restoring force for a perturbed fluid element, giving rise to a family of oscillation modes called $g$ modes \citep{1992ApJ...395..240R}.

In the following, the binary orbit is taken to be quasi-circular, in the sense that the inspiral evolves much more slowly than the orbital motion, namely $|\dot{D}|/D\ll\Omega_\text{orb}$ (with $\Omega_\text{orb}$ denoting the orbital frequency and the dot indicating differentiation with respect to time, as usual). To leading order, the orbital separation evolves due to the emission of gravitational waves as
\begin{equation}
    \dot{D}=-\frac{64G^3}{5c^5}\frac{MM'(M+M')}{D^3}
    \label{eq:OrbitalSeparationEvolution}
\end{equation}
($c$ being the speed of light), whereas for the orbital phase $\psi$ we have
\begin{equation}
    \dot{\psi}=\Omega_\text{orb}=\sqrt{\frac{G(M+M')}{D^3}},
    \label{eq:OrbitalPhaseEvolution}
\end{equation}
which, due to the approximation above, simply leads to $\psi\approx \Omega_\text{orb}t$.

For simplicity, we will assume that the orbital angular momentum is aligned with the rotation axis of the primary. In general, this need not be the case and an appropriate conversion between the two reference frames is required (\textit{e.g.}, see \citealt{1999MNRAS.308..153H, 2019PhRvD.100f3016P}). We also choose to work in the frame co-rotating with the primary. In order to re-express the tidal potential \eqref{eq:TidalField} in this frame, it simply suffices to make the replacement $\psi \rightarrow \psi - \Omega t$, keeping in mind that now the spherical harmonic $Y_l^m$ is expressed in the rotating frame. Hence, the time dependence of the tidal potential in this frame becomes $e^{-im[\psi(t)-\Omega t]}\approx e^{-im\bar{\Omega} t}$, where $\bar{\Omega}=\Omega_\text{orb} - \Omega$. Naturally, if the star is not rotating, the two frames coincide and $\bar{\Omega} = \Omega_\text{orb}$.

Ultimately, we are looking for a solution to \cref{eq:PerturbedEuler} that describes the full response of the star to the influence of the tidal field of its companion. One way to do this is to make use of the complete set of oscillation modes of the star.


\section{Normal modes of oscillation} \label{sec:Normal modes of oscillation}

A mode solution
\begin{equation}
    \xi^i(t, x^i) = \xi_\alpha^i(x^i) e^{i \omega_\alpha t},
    \label{eq:Mode}
\end{equation}
with frequency $\omega_\alpha$, is a distinct physical solution to the homogeneous equation of motion \eqref{eq:PerturbedEuler}, $\mathcal{E}_i = 0$, accompanied by the appropriate boundary conditions. As we assume a stable equilibrium with no dissipative processes, all mode frequencies are real. It follows that a mode solution satisfies the quadratic eigenvalue equation
\begin{equation}
    (- \omega_\alpha^2 g_{i j} + i \omega_\alpha B_{i j} 
    + C_{i j}) \xi_\alpha^j 
        = 0
    \label{eq:QuadraticEigenvalue}
\end{equation}
for the mode frequencies $\{ \omega_\alpha \}$ and eigenfunctions $\{ \xi_\alpha^i \}$. It is worth noting that both the set $(\omega_\alpha, \xi_\alpha^i)$ and the corresponding conjugate set $(- \omega_\alpha, \xi_\alpha^{i *})$ are solutions to \cref{eq:QuadraticEigenvalue}. We need to pay attention to this later, when the tidal displacement is decomposed in terms of the oscillation modes.

A key issue involves distinguishing the distinct mode solutions. This is resolved by making use of the formalism developed in \citet{1978ApJ...221..937F}. First, we need to introduce the inner product
\begin{equation}
    \langle \eta^i, \xi_i \rangle 
        = \int \eta^{i *} \xi_i \rho \dd V
    \label{eq:InnerProduct}
\end{equation}
of two complex solutions $\eta^i$ and $\xi^i$ to $\mathcal{E}_i = 0$. (Component indices are henceforth suppressed, for brevity.) With respect to this inner product, the operator $C$ is Hermitian, while $B$ is anti-Hermitian. Next, we define the \textit{symplectic product} of two solutions by
\begin{equation}
    W(\eta, \xi) 
        \equiv 
            \left\langle \eta, \partial_t \xi + \frac{1}{2} B \xi \right\rangle
            - 
            \left\langle \partial_t \eta + \frac{1}{2} B \eta, \xi \right\rangle,
\end{equation}
which can be shown to be conserved, so we have $\dd W/\dd t = 0$. This is particularly useful, because it implies that \textit{all} physical mode solutions must be orthogonal to each other \textit{with respect to $W$}. Provided two mode solutions of the form \eqref{eq:Mode}, we find
\begin{equation}
    W(\eta, \xi) 
        = \left( \left\langle \xi_\beta, 
            i \omega_\alpha \xi_\alpha + \frac{1}{2} B \xi_\alpha \right\rangle 
            - \left\langle i \omega_\beta \xi_\beta + \frac{1}{2} B \xi_\beta, 
            \xi_\alpha \right\rangle \right) e^{i (\omega_\alpha - \omega_\beta) t}
        \equiv W(\xi_\beta, \xi_\alpha) e^{i (\omega_\alpha - \omega_\beta) t},
\end{equation}
where we have defined the symplectic product of two eigenfunctions, $W(\xi_\beta, \xi_\alpha)$. Assuming no degeneracy, namely that $\omega_\alpha \neq \omega_\beta$ for $\alpha \neq \beta$, we therefore must have%
\footnote{If degeneracies are present, $W(\xi_\beta, \xi_\alpha)$ needs to be diagonalised within each degenerate subspace, namely within submatrices populated by modes belonging to the same eigenfrequency. The basis that diagonalises each degenerate subspace can be obtained as an appropriate linear combination of the vectors belonging to that degenerate subspace. Then, \cref{eq:SymplecticProduct} would also be satisfied for mode pairs from within the same degenerate subspace \citep{2001PhRvD..65b4001S}.}
\begin{equation}
    W(\xi_\beta, \xi_\alpha) 
        = i (\omega_\alpha + \omega_\beta) \langle \xi_\beta, \xi_\alpha \rangle 
        + \langle \xi_\beta, B \xi_\alpha \rangle
        = 0 .
    \label{eq:SymplecticProduct}
\end{equation}
Hence, we obtain the orthogonality condition
\begin{equation}
    (\omega_\alpha + \omega_\beta) \langle \xi_\beta, \xi_\alpha \rangle 
    - \langle \xi_\beta, i B \xi_\alpha \rangle 
        = \mathcal{B}_\alpha \delta_{\alpha \beta},
    \label{eq:RotatingOrthogonalityCondition}
\end{equation}
which, in the non-rotating limit, simply becomes
\begin{equation}
    \langle \xi_\beta, \xi_\alpha \rangle 
        = \mathcal{A}_\alpha^2 \delta_{\alpha \beta}
    \quad \text{as} \quad \Omega \to 0,
    \label{eq:NonRotatingOrthogonalityCondition}
\end{equation}
where $\mathcal{A}_\alpha^2$ and $\mathcal{B}_\alpha$ are (real) constants, determined by the chosen mode normalisation.

At this point, it is worth noting that the oscillation modes of a rotating star are not expected to satisfy \cref{eq:NonRotatingOrthogonalityCondition}. This is exactly the reason that the symplectic product was required in the first place; the modes obey the alternative orthogonality condition \eqref{eq:RotatingOrthogonalityCondition}, but are not orthogonal, in the conventional sense, with respect to the inner product \eqref{eq:InnerProduct}. This subtlety, which is often overlooked in the literature \citep[e.g., ][]{1999MNRAS.308..153H}, is very important as it modifies the mode expansion, here applied to the Love number calculation but generally used for a range of problems (see the discussion in \citealt{2001PhRvD..65b4001S}). As demonstrated later, one needs to pay attention to the specific notion of orthogonality in the rotating case in order to arrive at a mode-sum representation where the equations for the individual mode amplitudes decouple.

From \cref{eq:QuadraticEigenvalue}, after setting $B=0$, it becomes apparent that, since the operator $C$ is Hermitian, the modes of a non-rotating star form a complete basis, obeying the orthogonality condition \eqref{eq:NonRotatingOrthogonalityCondition}. The same cannot be said, however, for the general case of rotating stars, where the mode solutions $\{\xi_\alpha\}$ are eigenvectors of a non-Hermitian operator and, hence, need not form a complete basis. This is not immediately evident from \cref{eq:QuadraticEigenvalue}, which needs to be expressed as a standard eigenvalue equation with $\omega_\alpha$ as the eigenvalue.%
\footnote{This procedure, first presented in \citet{1979RSPSA.368..389D} and revisited in \citet{2001PhRvD..65b4001S}, involves re-writing the homogeneous equation of motion \eqref{eq:PerturbedEuler}, $\mathcal{E}_i=0$, as a system of first-order equations and then deriving an eigenvalue equation for eigenvector pairs $(\xi,\partial_t\xi)$ with $\omega$ as the eigenvalue.}
In this case, a way to obtain a complete basis is to supplement each eigenvector with a suitable number of extra vectors, called \textit{Jordan chain vectors}. However, a consistent treatment involving Jordan chains would complicate the formalism considerably and the benefits have not been rigorously explored. As this is beyond the scope of the current study, following \citet{2001PhRvD..65b4001S}, Jordan chain vectors will be neglected altogether and the mode solutions $\{\xi_\alpha\}$ to \cref{eq:QuadraticEigenvalue} will be treated as complete.


\section{Mode-sum representation} \label{sec:Mode-sum representation}

As advertised, we will express the tidal response of the primary star due to the presence of the secondary in terms of the complete basis given by the star's modes of oscillation. The strategy is common but, as we need to explore a number of subtle points, it still makes sense to tread carefully.

First, let us consider the nature of the required mode solutions. Given an axisymmetric background---which is clearly the case for both non-rotating and rotating stars---each mode $\xi_\alpha$ has a harmonic dependence in the azimuthal direction and is thus identified with a single azimuthal order $m$, \textit{i.e.}, $\xi_\alpha^i \propto e^{i (\omega_\alpha t + m \phi)}$. For non-rotating stars, due to the spherical symmetry of the background, each mode can be described by a single spherical harmonic $Y_l^m$ and, hence, corresponds to a single $l$. This implies that the eigenfrequencies $\omega_\alpha$ are degenerate with respect to $m$; for a fixed value of $l$, there are $(2l+1)$ modes with the same frequency $\omega_\alpha$. When rotation is introduced, this degeneracy is lifted by splitting the modes belonging to different values of $m$. Furthermore, modes can no longer be assigned a single degree $l$, due to induced couplings among different multipoles. We will return to this issue in \cref{subsec:Rotating stars}. The modes can be physically distinguished by noting that, for solutions with $m\neq 0$, the oscillation pattern propagates in the azimuthal direction with a phase velocity $-\omega_\alpha/m$. Given the Ansatz \eqref{eq:Mode} and for $\omega_\alpha > 0$, modes with $m<0$ travel along the direction of rotation (\textit{prograde} modes), whereas modes with $m>0$ travel in the opposite direction (\textit{retrograde} modes).

In order to establish the logic, and provide useful comparison, it is natural to first consider the non-rotating case.


\subsection{Non-rotating stars} \label{subsec:Non-rotating stars}

The non-rotating problem was recently explored in \citet{2020PhRvD.101h3001A}. In that case, the mode eigenvectors $\{ \xi_\alpha \}$ are complex solutions to \cref{eq:QuadraticEigenvalue} with $B_i^{\hphantom{i} j} = 0$, namely
\begin{equation}
    (- \omega_\alpha^2 g_{i j} + C_{i j}) \xi_\alpha^j = 0,
    \label{eq:NonRotatingQuadraticEigenvalue}
\end{equation}
with (real) eigenvalues $\{ \omega_\alpha^2 \}$. The completeness of the modes implies that a generic Lagrangian perturbation can be decomposed \citep[e.g., see][]{2001PhRvD..65b4001S} as
\begin{equation}
    \xi^i(t, x^i) = \sum_\alpha a_\alpha(t) \xi_\alpha^i(x^i),
    \label{eq:NonRotatingDecomposition}
\end{equation}
where the mode amplitude $a_\alpha$ is formally defined by
\begin{equation}
    a_\alpha = \frac{1}{\mathcal{A}_\alpha^2} \langle \xi_\alpha, \xi \rangle.
\end{equation}

This is important. Since the eigenvalue is $\omega_\alpha^2$, the mode solutions of a non-rotating star are independent of the sign of the mode frequency. Also, given the $e^{im\phi}$ azimuthal dependence, it is easy to show that a (real-frequency) mode solution and its complex conjugate are orthogonal (as required in order for the mode sum to be able to represent a real-valued function). Hence, in order to obtain a complete basis we can restrict ourselves to solutions with $\omega_\alpha \ge0$. This is the convention we adopt in the following.%
\footnote{Alternatively, one may instead fix the sign of the mode azimuthal order, $m>0$, and allow for both positive and negative frequencies \citep{1999MNRAS.308..153H,2006PhRvD..74b4007L,2017PhRvD..96h3005X,2021PSJ.....2..122L,2022ApJ...925..124D}. However, this convention is slightly confusing, because one has to pay careful attention to the complex conjugates in order to ensure that all distinct mode solutions are accounted for in the mode sum. \label{foot:FrequencySignConvention}}
To be specific, one can show that, despite working with complex eigenfunctions as the basis vectors in the mode expansion~\eqref{eq:NonRotatingDecomposition}, the physical perturbation $\xi^i$ is manifestly real. This is most easily seen by considering mode pairs $(m, - m)$ with the same eigenfrequency $\omega_\alpha$. These pairs combine in the mode summation to exactly cancel the imaginary parts. That this is the case should not be surprising since the field~\eqref{eq:TidalField} that sources the perturbation is real. Furthermore, when we proceed to the rotating case we need to keep in mind that each of the non-rotating modes split in two. We need to pay attention to the associated change in the symmetry of the mode frequencies later.

The perturbation $\xi^i$ is sourced by the tidal field of the companion. Inserting the formal expansion \eqref{eq:NonRotatingDecomposition} into the perturbed Euler equation \eqref{eq:PerturbedEuler}, and making use of the eigenvalue equation \eqref{eq:NonRotatingQuadraticEigenvalue}, leads to
\begin{equation}
    \ddot{a}_\alpha + \omega_\alpha^2 a_\alpha 
        = - \frac{1}{\mathcal{A}_\alpha^2} 
            \langle \xi_\alpha, \nabla \chi \rangle.
    \label{eq:NonRotatingDynamical}
\end{equation}
That is, we obtain the equation of motion for the mode amplitudes (simply a set of forced oscillators). It is natural to define the \textit{overlap integral}%
\footnote{Note that the definition of the overlap integral used here is more general than that used in some studies (\textit{e.g.}, \citealt{1977ApJ...213..183P}, \citealt{1999MNRAS.308..153H}), where it is instead defined as
\begin{equation*}
    I_\alpha = \left\langle\xi_\alpha,\nabla\left(r^l Y_l^m\right)\right\rangle
             = \int\delta\rho_\alpha^* r^l Y_l^m\dd V.
\end{equation*}
These are the mass multipole moments, which, after replacing $\delta\rho_\alpha$, take the form of Eq.~\eqref{eq:NonRotatingMassMultipoleMoment}.}
\begin{equation}
    Q_\alpha \equiv - \langle \xi_\alpha, \nabla \chi \rangle 
        = - \int \xi_\alpha^{i *} \nabla_i \chi \rho \dd V 
        = \int \nabla_i (\rho \xi_\alpha^{i *}) \chi \dd V 
        = - \int \delta \rho_\alpha^* \chi \dd V,
\end{equation}
arrived at through integration by parts, where we have identified the density perturbation $\delta \rho_\alpha$ associated with the mode by making use of the perturbed continuity equation~\eqref{eq:PerturbedContinuity}. Note that this definition of $Q_\alpha$ inherits the time dependence from $\chi$---we are not working in the frequency domain (in contrast to most previous work on this problem).

In the following, recall that we use a condensed notation where the index $\alpha$ carries information on the specific mode and therefore implicitly provides the $(l, m)$ harmonic dependence. Thus, the perturbed density eigenfunction is expanded as
\begin{equation}
    \delta \rho_\alpha (x^i) = \delta \rho_\alpha(r) Y_l^m(\theta, \phi).
\end{equation}
Then, the overlap integral is simply written as
\begin{equation}
    Q_\alpha = K_{l m} I_\alpha e^{- i m \Omega_\text{orb} t},
\end{equation}
where $K_{l m} \equiv G M' W_{l m} / D^{l + 1}$ and the \textit{mass-multipole moment} for each mode is defined by 
\begin{equation}
    I_\alpha \equiv \int_0^R \delta \rho_\alpha(r) r^{l + 2} \dd r,
    \label{eq:NonRotatingMassMultipoleMoment}
\end{equation}
noting that $\delta\rho_\alpha(r)$ and, thus, $I_\alpha$ are real-valued (see \cref{subsec:The multipole expansion approach} for the relevant arguments).

Returning to the equation of motion for the amplitudes \eqref{eq:NonRotatingDynamical}, each mode is driven by the tidal field at a frequency $m\Omega_\text{orb}$. As long as the orbital motion does not resonantly excite the mode, the amplitude equation of motion therefore gives%
\footnote{\Cref{eq:NonRotatingAmplitudeSolution} fails to describe the mode amplitude near the mode resonance, as it diverges. An analytical expression can be nevertheless obtained near the resonance, using the stationary phase approximation (\textit{e.g.}, see \citealt{2016PhRvD..94j4028S}).}
\begin{equation}
    a_\alpha(t) 
        = \frac{K_{l m} I_\alpha}{\mathcal{A}_\alpha^2 
            [\omega_\alpha^2 - (m \Omega_\text{orb})^2]}
            e^{- i m \Omega_\text{orb} t}.
    \label{eq:NonRotatingAmplitudeSolution}
\end{equation}
In essence, each mode $\alpha$ carries an $(l, m)$ dependence and will be excited by the corresponding $(l, m)$ component of the tidal field $\chi$. However, note that the tidal forcing is only non-zero for even $l + m$, so modes with odd $l + m$ will not be excited [see \cref{eq:Wlm}].

We can now proceed with the calculation of the tidal Love number $k_{lm}$. The Love number quantifies the multipolar response of a star to a given tidal field and is defined as
\begin{equation}
    \delta \Phi_{l m} = 2 k_{l m} \chi_{l m},
\end{equation}
evaluated at the surface the star.%
\footnote{Generally, the name \textit{Love number} is used for the tidal deformations due to a static external field. Here, we will say \textit{static} Love number to mean this, and just Love number or \textit{effective} Love number for the general case of a time-varying tidal field, representing the \textit{dynamical tide}.}
Here, the tidally-induced perturbation of the gravitational potential has been decomposed in spherical harmonics as
\begin{equation}
    \delta \Phi(t, x^i) 
        = \sum_{l, m} \delta \Phi_{l m}(t, r) Y_l^m(\theta, \phi),
\end{equation}
while the $(l,m)$ component of the tidal potential is simply
\begin{equation}
    \chi_{l m}(t, r) 
        = - \frac{G M' W_{l m} r^l}{D^{l + 1}} e^{- i m \Omega_\text{orb} t} 
        = - K_{l m} r^l e^{- i m \Omega_\text{orb} t}.
\end{equation}

In the vacuum exterior, the perturbed gravitational potential of the star is related to the multipole moments that characterise the departure from the equilibrium shape. Thus, a given mode sources
\begin{equation}
    \delta \Phi_\alpha(R) = - \frac{4 \pi G}{(2 l + 1) R^{l + 1}} I_\alpha
    \label{eq:GravitationalPotentialMultipoleMomentRelation}
\end{equation}
at the stellar surface. (Note that we have adopted a different sign convention for the multipole moments to that of \citealt{2020PhRvD.101h3001A}.) Using the fact that $\delta \Phi$ will inherit the decomposition~\eqref{eq:NonRotatingDecomposition} from $\xi^i$, through the linearised Poisson's equation~\eqref{eq:PerturbedPoissons}, and replacing the solution \eqref{eq:NonRotatingAmplitudeSolution} for the mode amplitudes, we have
\begin{equation}
    \delta \Phi(t, R, \theta, \phi) 
        = - 4 \pi G \sum_\alpha \frac{1}{(2 l + 1) R^{l + 1}} 
            \frac{I_\alpha^2}{\mathcal{A}_\alpha^2 
            [\omega_\alpha^2 - (m \Omega_\text{orb})^2]} 
            K_{l m} e^{- i m \Omega_\text{orb} t} Y_l^m(\theta, \phi).
\end{equation}
To identify the contribution of a certain harmonic $(l,m)$ of the gravitational potential, we introduce some further notation at this point. We use the label $\alpha'$ to denote the set of modes $\{ \alpha' \}$ belonging to a given $(l, m)$. Each subset of modes with fixed $(l,m)$ would then include modes from different classes and overtones (see \cref{sec:The slow-rotation approximation}). We thus have
\begin{equation}
    \delta \Phi(t, R, \theta, \phi) 
        = - 4 \pi G 
            \sum_l \frac{1}{(2 l + 1) R^{l + 1}} 
            \sum_m K_{l m} e^{- i m \Omega_\text{orb} t} 
            \sum_{\alpha'} \frac{I_{\alpha'}^2}{\mathcal{A}_{\alpha'}^2 
            [\omega_{\alpha'}^2 - (m \Omega_\text{orb})^2]} Y_l^m,
\end{equation}
and identify
\begin{equation}
    \delta \Phi_{l m}(t, R) 
        = - \frac{4 \pi G}{(2 l + 1) R^{l + 1}} K_{l m} 
            e^{- i m \Omega_\text{orb} t} 
            \sum_{\alpha'} \frac{I_{\alpha'}^2}{\mathcal{A}_{\alpha'}^2 
            [\omega_{\alpha'}^2 - (m \Omega_\text{orb})^2]}.
\end{equation}

Finally, we deduce that the effective Love number of a non-rotating star is
\begin{equation}
    k_{l m} 
        = \frac{2 \pi G}{(2 l + 1) R^{2 l + 1}} 
            \sum_{\alpha'} \frac{I_{\alpha'}^2}{\mathcal{A}_{\alpha'}^2 
            [\omega_{\alpha'}^2 - (m \Omega_\text{orb})^2]}.
    \label{eq:NonRotatingLove}
\end{equation}
The static Love number (obtained in the limit $\Omega_\text{orb}\to 0$) is simply given by
\begin{equation}
    k_l
        = \frac{2 \pi G}{(2 l + 1) R^{2 l + 1}} 
            \sum_{\alpha'} \frac{I_{\alpha'}^2}{\mathcal{A}_{\alpha'}^2 
            \omega_{\alpha'}^2}.
    \label{eq:NonRotatingStaticLove}
\end{equation}
As is evident from \cref{eq:NonRotatingStaticLove}, there is no dependence on the azimuthal order $m$ in the static limit. This is as expected, given that the mode pattern should only matter when the mode is resonantly excited by the tide, namely when $\omega_{\alpha'}\sim |m|\Omega_\text{orb}$. 

Note that \cref{eq:NonRotatingLove,eq:NonRotatingStaticLove} differ slightly from the corresponding expressions in \citet[Eqs.~(102) and (100), respectively]{2020PhRvD.101h3001A}. There, the gravitational potential $\delta\Phi$ was expressed in terms of the radial and horizontal components of the Lagrangian displacement $\xi^i$. However, one can demonstrate that the results are equivalent and converge to the same value for the Love number when the mode-sum is executed.


\subsection{Rotating stars} \label{subsec:Rotating stars}

We now turn our attention to the general case of rotating stars. This problem has been re-visited in recent years, particularly in the context of describing Jupiter's tidal Love numbers \citep{2021PSJ.....2..122L, 2022ApJ...925..124D}. However, as we will note, there are important details that still need to be satisfactorily explained. Specifically, the issue of the $l$-multipole coupling that exists with rotation, along with which modes sit in the sum, has not previously enjoyed the attention it deserves. Additionally, in the past, there has been confusion surrounding the orthogonality of the modes when rotation is accounted for \citep{1995ApJ...449..294K,1997ApJ...490..847L,1999MNRAS.308..153H}. In this section, we attempt to rectify these issues and make all relevant details clear. Indeed, this will be important if we want to connect observations of tides with the oscillation modes of the body.

Based on the discussion in \cref{sec:Normal modes of oscillation}, the mode solutions to the eigenvalue equation \eqref{eq:QuadraticEigenvalue} are treated as a complete set and, thus, it is possible to decompose a generic displacement in the spirit of \cref{eq:NonRotatingDecomposition}. For rotating stars, mode solutions for frequencies with the opposite sign remain degenerate, providing a similar eigenvector basis $\{ \xi_\alpha^i \}$ with linearly independent entries that satisfy the orthogonality relation~\eqref{eq:RotatingOrthogonalityCondition}. However, the decomposition~\eqref{eq:NonRotatingDecomposition} is no longer practical as the equations of motion for the amplitudes $\{ a_\alpha \}$ do not, in general, decouple \citep{2001PhRvD..65b4001S}. This can be illustrated by replacing the decomposition~\eqref{eq:NonRotatingDecomposition} in the perturbed Euler equation~\eqref{eq:PerturbedEuler}, which, making use of the eigenvalue equation~\eqref{eq:QuadraticEigenvalue} and the orthogonality condition~\eqref{eq:RotatingOrthogonalityCondition}, gives
\begin{equation}
    \mathcal{A}_\alpha^2\ddot{a}_\alpha+i\left(\mathcal{B}_\alpha-2\omega_\alpha \mathcal{A}_\alpha^2\right)\dot{a}_\alpha+\omega_\alpha\left(\mathcal{B}_\alpha-\omega_\alpha \mathcal{A}_\alpha^2\right)a_\alpha = Q_\alpha - \sum_{\beta\ne\alpha}\left[\ddot{a}_\beta-i\omega_\beta\dot{a}_\beta-i\omega_\alpha\left(\dot{a}_\beta-i\omega_\beta a_\beta\right)\right]\langle\xi_\alpha,\xi_\beta\rangle.
    \label{eq:RotatingDynamicalConfigurationSpace}
\end{equation}
Given that the inner product $\langle\xi_\alpha,\xi_\beta\rangle$ does not vanish for the modes of a rotating star, this inevitably leads to a coupling of all the equations of motion of different modes. 

This issue is resolved if one instead introduces a phase-space expansion of the form
\begin{equation}
    \begin{bmatrix}
        \xi^i(t, x^i) \\
        \partial_t \xi^i(t, x^i)
    \end{bmatrix}
        = \sum_A c_A(t) 
    \begin{bmatrix}
        \xi_A^i(x^i) \\
        i \omega_A \xi_A^i(x^i)
    \end{bmatrix},
    \label{eq:RotatingDecompositionGeneral}
\end{equation}
where the Lagrangian displacement and its time derivative are expanded simultaneously in terms of the basis vectors $(\xi_A^i, \, i\omega_A\xi_A^i)$. This strategy was originally developed by \citet{1979RSPSA.368..389D} and extensively explored by \citet{2001PhRvD..65b4001S} (see also the relevant discussion in \citealt{2017PhRvD..96l4008F}), where the interested reader may find useful details. Here we need to be careful, because, in the phase-space decomposition, the number of components $\{A\}$ in \cref{eq:RotatingDecompositionGeneral} is twice the number of $\{\alpha\}$ in \cref{eq:NonRotatingDecomposition}. Basically, the solutions $(\omega_\alpha,\xi_\alpha^i)$ and $(-\omega_\alpha,\xi_\alpha^{i*})$ are \emph{not} degenerate in the phase-space expansion~\eqref{eq:RotatingDecompositionGeneral}. However, as we are aiming to represent the real valued tidal response, we can pair up each mode with its conjugate, assigning the same label $\alpha$ to the pair of solutions, ensuring that $(\xi^i, \, \partial_t\xi^i)$ is manifestly real. Then, \cref{eq:RotatingDecompositionGeneral} becomes \citep{2001PhRvD..65b4001S}
\begin{equation}
    \begin{bmatrix}
        \xi^i(t, x^i) \\
        \partial_t \xi^i(t, x^i)
    \end{bmatrix}
        = \sum_\alpha \left\{ 
        c_\alpha(t) 
    \begin{bmatrix}
        \xi_\alpha^i(x^i) \\
        i \omega_\alpha \xi_\alpha^i(x^i)
    \end{bmatrix}
        + c_\alpha^*(t) 
    \begin{bmatrix}
        \xi_\alpha^{i *}(x^i) \\
        - i \omega_\alpha \xi_\alpha^{i *}(x^i)
    \end{bmatrix}
        \right\},
        \label{eq:RotatingDecomposition}
\end{equation}
noting that the mode frequencies $\omega_\alpha$ are taken to be positive (still applying the same convention as in the non-rotating case, following a similar line of reasoning; see \cref{subsec:Non-rotating stars}), with the amplitudes $c_\alpha$ formally defined by
\begin{equation}
    c_\alpha = \frac{1}{\mathcal{B}_\alpha}\langle\xi_\alpha,\omega_\alpha\xi-i\partial_t\xi-i B\xi)\rangle.
    \label{eq:RotatingAmplitudeCoefficients}
\end{equation}
Based on this, we can also derive the relation between the amplitudes $c_\alpha$, used in the phase-space expansion \eqref{eq:RotatingDecomposition}, and $a_\alpha$, from the expansion \eqref{eq:NonRotatingDecomposition}. Replacing \cref{eq:NonRotatingDecomposition} in \cref{eq:RotatingAmplitudeCoefficients}, we get
\begin{equation}
    c_\alpha = -\frac{i}{\mathcal{B}_\alpha}\big[\mathcal{A}_\alpha^2\dot{a}_\alpha+i\left(\mathcal{B}_\alpha-\omega_\alpha \mathcal{A}_\alpha^2\right)a_\alpha+\sum_{\beta\ne\alpha}\left(\dot{a}_\beta-i\omega_\beta a_\beta\right)\langle\xi_\alpha,\xi_\beta\rangle\big],
\end{equation}
where the coupling of the amplitudes corresponding to different modes is (again) evident.

Replacing the mode solutions $\{ A \}$ with $\{ \alpha \}$ in the phase-space expansion \eqref{eq:RotatingDecompositionGeneral} is not simply a choice. The identification of $\{ \alpha \}$, as representing all the \textit{distinct} mode solutions, is necessary in order to distinguish the contribution of each mode to the tidal response. The identification makes it clear precisely which modes are used in the sum and removes redundancy from the problem. In contrast, a  strategy based on using  $\{ A \}$ (as in \citealt{2021PSJ.....2..122L} and \citealt{2022ApJ...925..124D}) involves potential double-counting, so the final results have to be employed with some  level of care.

As in the non-rotating case, the time dependence of the perturbation $\xi^i$ is dictated by the tidal field. Inserting the decomposition~\eqref{eq:RotatingDecomposition} in the perturbed Euler equation~\eqref{eq:PerturbedEuler}, and making use of the eigenvalue equation~\eqref{eq:QuadraticEigenvalue} and the orthogonality condition~\eqref{eq:RotatingOrthogonalityCondition}, we arrive at the equation of motion for the amplitudes
\begin{equation}
    \dot{c}_\alpha - i \omega_\alpha c_\alpha 
        = \frac{i}{\mathcal{B}_\alpha} \langle \xi_\alpha, \nabla \chi \rangle
        = - \frac{i}{\mathcal{B}_\alpha} Q_\alpha.
    \label{eq:RotatingDynamical}
\end{equation}
It also follows that
\begin{equation}
    \dot{c}_\alpha^* + i \omega_\alpha c_\alpha^* 
        = \frac{i}{\mathcal{B}_\alpha} Q_\alpha^*.
\end{equation}

In general, the modes of a rotating star no longer correspond to a single $l$; they become coupled. The perturbed density associated with a mode can then be expressed as
\begin{equation}
    \delta \rho_\alpha = \sum_l \delta \rho_{\alpha l} Y_l^m, \label{eq:PerturbedDensityMultipoleExpansion}
\end{equation}
which leads to
\begin{equation}
    Q_\alpha = \sum_l K_{l m} I_{\alpha l} e^{- i m \bar{\Omega} t}.
\end{equation}
The time dependence of $Q_\alpha$ is again inherited from the tidal potential. We note that this coupling among different multipoles was missing from the analysis of \citet{2021PSJ.....2..122L}, although it was later rectified in \citet{2022ApJ...925..124D}.

Since we are working in the rotating frame, each mode is driven by the tidal field at a frequency $m\bar{\Omega}$. Solving \cref{eq:RotatingDynamical} away from resonance, we find
\begin{equation}
    c_\alpha(t) 
        = \frac{1}{\mathcal{B}_\alpha (\omega_\alpha + m \bar{\Omega})} 
            \sum_l K_{l m} I_{\alpha l} e^{- i m \bar{\Omega} t}.
    \label{eq:RotatingAmplitudeSolution}
\end{equation}

The perturbed gravitational potential can be expanded as
\begin{equation}
    \delta \Phi(t, x^i) 
        = \sum_\alpha \sum_l 
            [c_\alpha(t) \delta \Phi_{\alpha l}(r) Y_l^m(\theta, \phi) 
            + 
            c_\alpha^*(t) \delta \Phi_{\alpha l}^*(r) Y_l^{m *}(\theta, \phi)].
\end{equation}
In this expression, we sum over all the mode pairs $\{ \alpha \}$. Each $\alpha$ will be associated with a single $m$, but will couple different values of $l$. The sum over $l$ will pick up these coupled values of $l$ for each $\alpha$. At the surface of the star, replacing the solution for the amplitudes~\eqref{eq:RotatingAmplitudeSolution} and $\delta\Phi_{\alpha l}$ with \cref{eq:GravitationalPotentialMultipoleMomentRelation}, we have
\begin{equation}
    \delta \Phi(t, R, \theta, \phi) 
        = - 4 \pi G \sum_\alpha 
            \frac{1}{\mathcal{B}_\alpha (\omega_\alpha + m \bar{\Omega})} 
            \sum_l \frac{1}{(2 l + 1) R^{l + 1}} 
            \sum_{l'} K_{l' m} I_{\alpha l'} I_{\alpha l} 
            ( e^{- i m \bar{\Omega} t} Y_l^m 
            + e^{i m \bar{\Omega} t} Y_l^{m *} ).
\end{equation}
In the same vein as in the non-rotating case, we now introduce the label $\alpha'$ to represent all modes that have a given azimuthal order $m$. Hence, we obtain
\begin{equation}
    \delta \Phi(t, R, \theta, \phi) 
        = - 4 \pi G \sum_l 
            \frac{1}{(2 l + 1) R^{l + 1}} \sum_m \sum_{\alpha'} 
            \frac{1}{\mathcal{B}_{\alpha'} (\omega_{\alpha'} + m \bar{\Omega})} 
            \sum_{l'} K_{l' m} I_{\alpha' l'} I_{\alpha' l} 
            ( e^{- i m \bar{\Omega} t} Y_l^m 
            + e^{i m \bar{\Omega} t} Y_l^{m *} ).
\end{equation}
At this point, we have to be careful. We want to identify the contribution from each multipole to the perturbed potential. In order to do this, it is useful to note the identity $Y_l^{m *} = (- 1)^m Y_l^{- m}$. Next, we recall that we sum over all permissible values of $m$. Keeping this in mind, we can rewrite the second term by simply changing the sign of $m$. However, in doing so, we need to take care because it means that we tinker with the labelling of the modes. In order to remain mindful of this, we replace $\alpha'$ with $\beta'$ for the second set of modes. This has two advantages. First, it allows us to identify the overall contribution from a given spherical harmonic. Second, as we will see later, it provides an intuitive understanding of how the tidal response for a given multipole involves both the prograde and retrograde modes. Previous calculations do not show these two sets explicitly. We arrive at
\begin{equation}
    \delta \Phi(t, R, \theta, \phi) 
        = - 4 \pi G \sum_l \frac{1}{(2 l + 1) R^{l + 1}} \sum_m 
            e^{- i m \bar{\Omega} t} \sum_{l'} K_{l' m}
            \left[ \sum_{\alpha'} 
            \frac{I_{\alpha' l'} I_{\alpha' l}}
            {\mathcal{B}_{\alpha'} (\omega_{\alpha'} + m \bar{\Omega})} 
            +  \sum_{\beta'} 
            \frac{I_{\beta' l'} I_{\beta' l}}
            {\mathcal{B}_{\beta'} (\omega_{\beta'} - m \bar{\Omega})} \right]
            Y_l^m,
\end{equation}
where we have used the fact that $K_{l', - m} = (- 1)^m K_{l' m}$. In this expression, we see how the modes are paired. The first term corresponds to the mode $(\omega_{\alpha'}, m)$ and the second is its partner $(\omega_{\beta'}, - m)$ moving in the opposite direction. One of the modes is prograde and the other is retrograde (depending on the sign of $m$). Therefore, as before,
\begin{equation}
    \delta \Phi_{l m}(t, R) 
        = - \frac{4 \pi G}{(2 l + 1) R^{l + 1}} 
            e^{- i m \bar{\Omega} t} \sum_{l'} K_{l' m} \left[ \sum_{\alpha'} 
            \frac{I_{\alpha' l'} I_{\alpha' l}}
            {\mathcal{B}_{\alpha'} (\omega_{\alpha'} + m \bar{\Omega})} 
            + \sum_{\beta'} 
            \frac{I_{\beta' l'} I_{\beta' l}}
            {\mathcal{B}_{\beta'} (\omega_{\beta'} - m \bar{\Omega})} \right],
\end{equation}
and the effective Love number is
\begin{equation}
    k_{l m} 
        = \frac{2 \pi G}{(2 l + 1) R^{2 l + 1}} \frac{1}{K_{l m}} 
            \sum_{l'} K_{l' m} \left[ \sum_{\alpha'} 
            \frac{I_{\alpha' l'} I_{\alpha' l}}
            {\mathcal{B}_{\alpha'} (\omega_{\alpha'} + m \bar{\Omega})} 
            +  \sum_{\beta'} 
            \frac{I_{\beta' l'} I_{\beta' l}}
            {\mathcal{B}_{\beta'} (\omega_{\beta'} - m \bar{\Omega})} \right].
    \label{eq:RotatingLove}
\end{equation}

As a sanity check of the result, note that, in the non-rotating limit $(\Omega=0)$, each mode is characterised by a single degree $l'$, so the only term retained in the sum over $l'$ is the one for which $l'=l$. Furthermore, the mode pairs $(m, \, - m)$ have the same eigenfrequency $(\omega_{\alpha'} = \omega_{\beta'})$ and multipole moment $(I_{\alpha'}=I_{\beta'})$, whereas $\mathcal{B}_{\alpha'}\to 2\omega_{\alpha'}\mathcal{A}_{\alpha'}^2$. Thus, \cref{eq:RotatingLove} straightforwardly reduces to \cref{eq:NonRotatingLove}.

The expression that we have derived for the effective Love number~\eqref{eq:RotatingLove} has some subtle but important differences compared to the corresponding results from \citet{2021PSJ.....2..122L} and \citet{2022ApJ...925..124D}. First of all, we explicitly sum over the distinct mode solutions $\{ \alpha \}$, whereas \citet{2021PSJ.....2..122L} and \citet{2022ApJ...925..124D} implicitly sum over the larger set of solutions $\{ A \}$. This is not a critical issue---as long as you know what you are doing---but it obfuscates the fact that mode solutions from both sets $(\omega_\alpha,\xi_\alpha^i)$ and $(-\omega_\alpha,\xi_\alpha^{i*})$ are included in the Love number expansion. This distinction is necessary in order to unambiguously identify which modes contribute to the effective Love number for a specific multipole $(l,m)$ of the tidal potential, and explicitly demonstrate the presence of pro- and retrograde modes. Secondly, \citet{2021PSJ.....2..122L} neglects higher-order rotational effects and, therefore, does not include the $l$-coupling for the mode eigenfunctions that arises due to the Coriolis force. Our expression~\eqref{eq:RotatingLove} involves no such approximation and is thus true for arbitrary rates of rotation. The final difference lies in the convention for the modes; in particular, making the Ansatz $\xi_\alpha^i \propto e^{-i\omega_\alpha t}$ [which is the opposite sign convention from the one we use here; see \cref{eq:Mode}] and, more importantly, fixing the sign of $m$ (instead of that of $\omega_\alpha$) for describing the mode solutions (see Footnote~\ref{foot:FrequencySignConvention}). Then, the results rely on the judicious use of complex conjugates to arrive at the actual tidal response, whereas, in our case, the reality of the result is guaranteed by the chosen convention, making the strategy more transparent.


\section{The slow-rotation approximation} \label{sec:The slow-rotation approximation}

The derivation of the dynamical tide, leading to \cref{eq:RotatingLove} for the Love number, is valid for arbitrary rotation rates. However, solving the corresponding eigenvalue equation \eqref{eq:QuadraticEigenvalue} for a rotating star is not easy, due to the complications induced by rotation. As already mentioned, the different multipoles couple and the modes no longer have a simple spherical harmonic angular dependence. On top of that, the centrifugal acceleration (which enters at order $\Omega^2$) causes the background to deviate from spherical symmetry. In addition, at order $\Omega$, a new set of oscillation modes emerges due to the Coriolis force; the so-called inertial modes \citep{1978MNRAS.182..423P,1999ApJ...521..764L}. 

The general problem is complicated, but it is important to clarify how the mode-sum result  can be used. We can make general progress in this direction by considering two commonly adopted strategies. First, if we want to proceed analytically as far as possible (which is useful to gain insight), then it is natural to treat the rotation perturbatively, \textit{i.e.}, employing the \textit{slow-rotation approximation}, where the eigenfrequencies and the eigenfunctions are expanded with respect to $\Omega$ (see, for example, \citealt{1989nos..book.....U} and \citealt{2010aste.book.....A}). At first order in rotation, due to the Coriolis acceleration, the degeneracy with respect to the azimuthal order $m$ is lifted (see \cref{sec:Normal modes of oscillation}) and different multipoles couple, whereas inertial modes are also added to the mode spectrum. At second order, the centrifugal force deforms the star into an oblate spheroid, thus also affecting the equilibrium configuration, and additional couplings and corrections are introduced to the perturbations \citep{1981ApJ...244..299S}, which are further corrected at third order \citep{1998A&A...334..911S} and so on. An alternative to the perturbative strategy is to approach the problem via \textit{time evolutions of the perturbation equations} \citep{2002MNRAS.334..933J, 2009MNRAS.394..730P, 2009PhRvD..80f4026G, 2011PhRvL.107j1102G, 2011PhRvD..83f4031G, 2021FrASS...8..166K}. This neatly avoids the complications associated with the multipole coupling, but has the drawback that it is not so easy to reliably extract the mode eigenfunctions \citep{2004MNRAS.352.1089S}. This, in turn, means that the evaluation of various perturbative quantities, \textit{e.g.}, the overlap integrals, becomes less reliable. Hence, we will focus on the perturbative approach here.

Within the slow-rotation framework, we may also adopt different strategies. The first involves working out the rotational corrections to the eigenfunctions as a formal expansion in the modes of the non-rotating star. In this case, one has to include all solutions that formally belong to the zero-frequency subspace \citep{1978ApJ...221..937F,1999ApJ...521..764L,2001PhRvD..65b4001S}, which, once rotation is introduced, become the inertial modes, thus complicating the issue. The second approach uses a multipole expansion of the perturbed quantities and then solves directly for the rotational corrections. Each approach has its merits and we will find it useful to comment on both.


\subsection{The mode expansion approach} \label{subsec:The mode expansion approach}

To start with, we expand the eigenfrequencies and the eigenfunctions (in the rotating frame) as
\begin{align}
	\omega_\alpha &= \omega^{(0)}_\alpha +\omega^{(1)}_\alpha\left(\Omega\right) +\mathcal{O}\left(\Omega^2\right), \\
	\xi_\alpha &= \xi^{(0)}_\alpha +\xi^{(1)}_\alpha\left(\Omega\right) +\mathcal{O}\left(\Omega^2\right),
\end{align}
where the superscript $(0)$ corresponds to the solution in the non-rotating limit, obtained from \cref{eq:NonRotatingQuadraticEigenvalue}, and the rest of the terms are rotational corrections. (Again, for brevity, we suppress component indices in this section.) As our main interest is in illustrating the principles involved, we  only consider first-order corrections here. This is convenient, because we know that the equilibrium quantities, \textit{i.e.}, density, pressure, and gravitational potential, are only affected by $\Omega$ at second order. Hence, for the operator $C$ we have $C=C^{(0)}+\mathcal{O}\left(\Omega^2\right)$, whereas the operator $B$ is, by definition, $B=B^{(1)}\left(\Omega\right)$. Then, the eigenvalue equation \eqref{eq:QuadraticEigenvalue} at first order in $\Omega$ is
\begin{equation}
    -\omega_\alpha^{(0)2}\xi_\alpha^{(1)}+C^{(0)}\xi_\alpha^{(1)}-2\omega_\alpha^{(0)}\omega_\alpha^{(1)}\xi_\alpha^{(0)}+i\omega_\alpha^{(0)}B^{(1)}\xi_\alpha^{(0)} = 0. \label{eq:QuadraticEigenvalueFirstOrder}
\end{equation}

Next, we expand the rotational corrections in terms of the modes of the non-rotating star (in the spirit of the mode-sum representation for the tidal response). We thus have
\begin{equation}
	\xi_\alpha^{(1)} = \sum_\beta c_{\alpha\beta}^{(1)}\xi_\beta^{(0)}, \label{eq:EigenfunctionFirstOrderCorrectionExpansion}
\end{equation}
where $c_{\alpha\beta}^{(1)}$ are the first-order correction coefficients. Replacing \cref{eq:EigenfunctionFirstOrderCorrectionExpansion} into the first-order eigenvalue equation \eqref{eq:QuadraticEigenvalueFirstOrder} and using \cref{eq:NonRotatingQuadraticEigenvalue}, we obtain
\begin{equation}
	\sum_\beta c_{\alpha\beta}^{(1)}\left(\omega_\beta^{(0)2}-\omega_\alpha^{(0)2}\right)\xi_\beta^{(0)}-2\omega_\alpha^{(0)}\omega_\alpha^{(1)}\xi_\alpha^{(0)}+i\omega_\alpha^{(0)}B^{(1)}\xi_\alpha^{(0)} = 0.
\end{equation}
Taking the inner product with $\xi_\beta^{(0)}$ and using the mode orthogonality condition \eqref{eq:NonRotatingOrthogonalityCondition}, we have
\begin{equation}
	c_{\alpha\beta}^{(1)}\left(\omega_\beta^{(0)2}-\omega_\alpha^{(0)2}\right)\mathcal{A}_\beta^{(0)2}-2\omega_\alpha^{(0)}\omega_\alpha^{(1)}\mathcal{A}_\beta^{(0)2}\delta_{\alpha\beta}+i\omega_\alpha^{(0)}\left\langle\xi_\beta^{(0)},B^{(1)}\xi_\alpha^{(0)}\right\rangle = 0, \label{eq:QuadraticEigenvalueFirstOrderExpanded}
\end{equation}
which, for $\beta=\alpha$ (and $\omega_\alpha^{(0)}\ne 0)$ gives the first-order correction to the eigenfrequency as
\begin{equation}
	\omega_\alpha^{(1)}=\frac{i\left\langle\xi_\alpha^{(0)},B^{(1)}\xi_\alpha^{(0)}\right\rangle}{2\mathcal{A}_\alpha^{(0)2}}, \label{eq:EigenfrequencyFirstOrderCorrection}
\end{equation}
whereas, for $\beta\ne\alpha$, we get
\begin{equation}
	c_{\alpha\beta}^{(1)}=\frac{i\omega_\alpha^{(0)}\left\langle\xi_\beta^{(0)},B^{(1)}\xi_\alpha^{(0)}\right\rangle}{\mathcal{A}_\beta^{(0)2}\left(\omega_\alpha^{(0)2}-\omega_\beta^{(0)2}\right)}. \label{eq:EigenfunctionFirstOrderCorrectionCoefficients}
\end{equation}

Before we proceed, we note that the arguments leading to \cref{eq:EigenfrequencyFirstOrderCorrection,eq:EigenfunctionFirstOrderCorrectionCoefficients} are valid only if the mode eigenfrequencies in the non-rotating case are non-degenerate, namely as long as $\omega_\beta^{(0)}\ne\omega_\alpha^{(0)}$ for all $\beta\ne\alpha$. However, we already know that this is not the case, because, for a mode with a fixed value of $l$, there is a degeneracy with respect to the azimuthal order $m$, with $(2l+1)$ modes corresponding to the same eigenfrequency. For mode pairs within the same degenerate subspace, \cref{eq:QuadraticEigenvalueFirstOrderExpanded} would imply that 
\begin{equation*}
    B_{\beta\alpha}^{(1)}\equiv\left\langle\xi_\beta^{(0)},B^{(1)}\xi_\alpha^{(0)}\right\rangle = 0,
\end{equation*}
which is not necessarily true; the matrix $B_{\beta\alpha}^{(1)}$ need not be diagonal within each degenerate subspace of modes. Hence, in general, one needs to switch to a basis in which this matrix is diagonal within each degenerate subspace. Then, one may repeat the arguments above using the new basis. In this case, \cref{eq:EigenfrequencyFirstOrderCorrection,eq:EigenfunctionFirstOrderCorrectionCoefficients} would have the same form, with the eigenfunction set $\{\xi_\alpha^{(0)}\}$ replaced with the new basis. The process for obtaining this basis can be found in many textbooks (\textit{e.g.}, \citealt{1970mmp..book.....M}), so will not be reproduced here.

Having worked out the rotational corrections to the eigenfunctions, one may calculate the corresponding corrections to any perturbed quantity, using the equations of \cref{sec:Tidal deformations}; \textit{e.g.}, for $\delta\rho^{(1)}$ one needs the perturbed continuity equation \eqref{eq:PerturbedContinuity} at first order in $\Omega$, namely $\delta\rho^{(1)}=-\nabla^i(\rho\xi^{(1)}_i)$. The obvious caveat of this approach is that it requires prior knowledge of as many modes of the non-rotating star as possible (in principle, all of them!), to be included in the  expansion \eqref{eq:EigenfunctionFirstOrderCorrectionExpansion}. Even though, practically, a few modes may be enough for the mode sum to sufficiently converge, the modes that contribute the most to the correction would have to be identified, so further theoretical reasoning would be necessary if one wants to be sure of the result.

This problem can in principle be avoided by solving directly for the first-order corrections to the various quantities, using the approach described next.


\subsection{The multipole expansion approach} \label{subsec:The multipole expansion approach}

As an alternative to expanding the eigenfunction corrections in terms of the modes of the non-rotating star, we may use a multipole expansion \citep[e.g., see][]{1981ApJ...244..299S}, in the spirit of \cref{eq:PerturbedDensityMultipoleExpansion}. Thus, scalar perturbations (here using the density perturbation as an example) are expanded as
\begin{equation}
    \delta\rho_\alpha = \sum_l \delta\rho_{\alpha l} Y_l^m,
\end{equation}
while the displacement vector is expanded in vector spherical harmonics as
\begin{equation}
    \xi_\alpha^i = \sum_l \left[
        \frac{W_{\alpha l}}{r}Y_l^m\nabla^i r
            + V_{\alpha l}\nabla^i Y_l^m
            -i U_{\alpha l}\epsilon^{ijk}\nabla_j Y_l^m\nabla_k r
        \right],
\end{equation}
or, in terms of components,
%
%
\begin{equation}
    \xi_\alpha^r = \frac{1}{r} \sum_l W_{\alpha l}Y_l^m,
    \quad
    \xi_\alpha^\theta = \frac{1}{r^2} \sum_l \left( V_{\alpha l} \partial_\theta Y_l^m - i U_{\alpha l}\frac{\partial_\phi Y_l^m}{\sin\theta} \right),
    \quad
    \xi_\alpha^\phi = \frac{1}{r^2\sin^2\theta} \sum_l \left( V_{\alpha l} \partial_\phi Y_l^m + i U_{\alpha l} \sin\theta\partial_\theta Y_l^m \right).
\end{equation}

Based on this decomposition, oscillation modes are often distinguished as \textit{polar} ($U_{\alpha l}=0$ as $\Omega\to 0$) or \textit{axial} ($W_{\alpha l}=V_{\alpha l}=0$ as $\Omega\to 0$). In the simple case of a fluid star, there are three families of polar modes: $f$ (fundamental) modes, $p$ (pressure) modes, and $g$ (gravity) modes, with the latter being associated with the presence of buoyancy in the star (due to, \textit{e.g.}, composition gradients; see discussion in \cref{sec:Tidal deformations}). The $p$ and $g$ modes are often also ordered as increasingly high and low frequency overtones, respectively. Axial modes become oscillatory when rotation is switched on and are called $r$ (Rossby) modes \citep{1978MNRAS.182..423P}. In the special case of a star where there is no buoyancy (described by a barotropic equation of state and with the oscillations adjusting rapidly to the density of their surroundings), the $f$ and $p$ modes are supplemented by the \textit{inertial} modes \citep{1999ApJ...521..764L,1999PhRvD..59d4009L}, which have both polar and axial components and are generally not characterised by a single $l$. Like $r$ modes, the inertial modes acquire non-zero frequencies only in the presence of rotation.

As we are interested in the first-order in $\Omega$ contributions  to the Love number, it suffices to consider only polar modes in \cref{eq:RotatingLove}, because their mass multipole moments are non-vanishing already at zeroth order in $\Omega$. The mass multipole moments of $r$ modes and inertial modes come in at order $\Omega^2$ \citep{1999ApJ...521..764L}, which implies that they should produce fourth-order corrections to the Love number.

As already mentioned, scalar perturbations associated with polar modes are simply decomposed as $\delta\rho_\alpha^{(0)}(x^i)=\delta\rho_\alpha^{(0)}(r)Y_l^m(\theta,\phi)$. The corresponding polar mode eigenfunctions can be expressed as
\begin{equation}
    \xi_\alpha^{(0)i} 
        = \frac{W_\alpha^{(0)}}{r}Y_l^m \nabla^i r
            + V_\alpha^{(0)} \nabla^i Y_l^m.
    \label{eq:EigenfunctionPolar}
\end{equation}
%
%
%
%
Using the definitions \eqref{eq:OperatorB} and \eqref{eq:OperatorC} of the operators $B$ and $C$, we also obtain
\begin{gather}
    B^{(1)}_{r j}\xi^{(0)j}_\alpha = -2\Omega \frac{V^{(0)}_\alpha}{r}\partial_\phi Y_l^m,
    \;
    B^{(1)}_{\theta j}\xi^{(0)j}_\alpha = -2\Omega V^{(0)}_\alpha\cot\theta\partial_\phi Y_l^m,
    \;
    B^{(1)}_{\phi j}\xi^{(0)j}_\alpha = 2\Omega \sin\theta\left(V^{(0)}_\alpha\cos\theta\partial_\theta Y_l^m+W^{(0)}_\alpha\sin\theta Y_l^m\right),
    \label{eq:OperatorBPolar}
    \intertext{and}
    C^{(0)}_{ij}\xi^{(1)j}_\alpha =
        \frac{\nabla_i\delta p^{(1)}_\alpha}{\rho}-\frac{\nabla_i p}{\rho^2}\delta\rho^{(1)}_\alpha+\nabla_i\delta\Phi^{(1)}_\alpha
        = \nabla_i\left(\frac{\delta p^{(1)}_\alpha}{\rho}+\delta\Phi^{(1)}_\alpha\right)-\frac{p\Gamma_1}{\rho}A_i\nabla_j\xi^{(1)j}_\alpha.
    \label{eq:OperatorCPolarFirstOrder}
\end{gather}

Now, we can use the multipole expansions above to express first-order quantities. Replacing in the first-order eigenvalue equation \eqref{eq:QuadraticEigenvalueFirstOrder} and taking the radial component, we get
\begin{equation}
    \sum_{l'} \left(
        -\omega^{(0)2}_\alpha \frac{W^{(1)}_{\alpha l'}}{r}
            +\frac{\partial_r\delta p^{(1)}_{\alpha l'}}{\rho}-\frac{\partial_r p}{\rho^2}\delta\rho^{(1)}_{\alpha l'}+\partial_r\delta\Phi^{(1)}_{\alpha l'}
    \right) Y_{l'}^m
            = \frac{2\omega^{(0)}_\alpha}{r}\left(\omega^{(1)}_\alpha W^{(0)}_\alpha
            -m\Omega V^{(0)}_\alpha \right)Y_l^m.
\end{equation}
%
%
%
Taking the inner product with $Y_l^m$ and using the orthogonality of spherical harmonics, this becomes
\begin{equation}
    -\omega^{(0)2}_\alpha \frac{W^{(1)}_{\alpha l}}{r}
            +\frac{\partial_r\delta p^{(1)}_{\alpha l}}{\rho}-\frac{\partial_r p}{\rho^2}\delta\rho^{(1)}_{\alpha l} + \partial_r\delta\Phi^{(1)}_{\alpha l}
        = \frac{2\omega^{(0)}_\alpha}{r}\left(\omega^{(1)}_\alpha W^{(0)}_\alpha
            -m\Omega V^{(0)}_\alpha \right),
    \label{eq:QuadraticEigenvalueFirstOrderRadial}
\end{equation}
or, equivalently [see \cref{eq:OperatorCPolarFirstOrder}],
\begin{equation}
    -\omega^{(0)2}_\alpha \frac{W^{(1)}_{\alpha l}}{r}
        + \partial_r\left( \frac{\delta p^{(1)}_{\alpha l}}{\rho}+\delta\Phi^{(1)}_{\alpha l} \right)
        -\frac{p\Gamma_1}{\rho} \frac{A}{r^2}\left[ \partial_r\left(r W^{(1)}_{\alpha l} \right)-l(l+1)V^{(1)}_{\alpha l} \right]
    = \frac{2\omega^{(0)}_\alpha}{r}\left(\omega^{(1)}_\alpha W^{(0)}_\alpha
            -m\Omega V^{(0)}_\alpha \right),
    \label{eq:QuadraticEigenvalueFirstOrderRadial2}
\end{equation}
where $A$ is the (magnitude of the) Schwarzschild discriminant \eqref{eq:SchwarzschildDiscriminant} (which only has an $r$ component at first order).

In a similar manner, we can take the inner product of \cref{eq:QuadraticEigenvalueFirstOrder} with $\nabla^i Y_l^m$, in order to obtain its horizontal polar component. This gives
\begin{equation}
    -\omega^{(0)2}_\alpha V^{(1)}_{\alpha l}
        + \frac{\delta p^{(1)}_{\alpha l}}{\rho}+\delta\Phi^{(1)}_{\alpha l}
        = 2\omega^{(0)}_\alpha\left[
            \omega^{(1)}_\alpha V^{(0)}_\alpha - \frac{m\Omega}{l(l+1)}\left(W^{(0)}_\alpha+V^{(0)}_\alpha\right)
        \right].
    \label{eq:QuadraticEigenvalueFirstOrderHorizontalPolar}
\end{equation}

\Cref{eq:QuadraticEigenvalueFirstOrderRadial,eq:QuadraticEigenvalueFirstOrderRadial2,eq:QuadraticEigenvalueFirstOrderHorizontalPolar} allude to the fact that the first-order polar corrections, $W^{(1)}_\alpha$ and $V^{(1)}_\alpha$, as well as the corrections to scalar perturbations, are sourced by a single multipole and, thus, pick up the same spherical harmonic dependence as their zeroth-order counterparts. Namely, corrections to scalar perturbations are simply decomposed as $\delta\rho_\alpha^{(1)}(x^i)=\delta\rho_{\alpha l}^{(1)}(r)Y_l^m(\theta,\phi)$, whereas the correction to the eigenfunction is
\begin{equation}
    \xi_\alpha^{(1)i} 
        = \frac{W_{\alpha l}^{(1)}}{r}Y_l^m \nabla^i r
            + V_{\alpha l}^{(1)} \nabla^i Y_l^m
            - i\sum_{l'} U^{(1)}_{\alpha l'} \epsilon^{ijk}\nabla_j Y_{l'}^m\nabla_k r
\end{equation}
(the axial components of the eigenfunction correction are yet to be determined).

The fact that first-order scalar perturbations are described by a single multipole can also be seen from \cref{eq:EigenfunctionFirstOrderCorrectionCoefficients}. From the perturbed continuity equation \eqref{eq:PerturbedContinuity}, the perturbed Poisson's equation \eqref{eq:PerturbedPoissons}, and the equation of state for the perturbations \eqref{eq:PerturbedEoS2} one can show that only polar pieces contribute to the first-order corrections of scalar quantities. Therefore, to obtain these corrections, only polar mode components are relevant in the expansion \eqref{eq:EigenfunctionFirstOrderCorrectionExpansion}. Hence, replacing the polar mode eigenfunction \eqref{eq:EigenfunctionPolar} for the mode $\xi^{(0)}_\beta$ in \cref{eq:EigenfunctionFirstOrderCorrectionCoefficients}, we obtain the corresponding first-order correction coefficients as
\begin{equation}
    c_{\alpha\beta}^{(1)} =
        \delta_{l_\alpha l_\beta}\delta_{m_\alpha m_\beta}\frac{2m_\alpha\Omega\omega_\alpha^{(0)}}{\mathcal{A}_\beta^{(0)2}\left(\omega_\alpha^{(0)2}-\omega_\beta^{(0)2}\right)}
        \int_0^R\left(W_\alpha^{(0)}V_\beta^{(0)}+V_\alpha^{(0)}W_\beta^{(0)}+V_\alpha^{(0)}V_\beta^{(0)}\right)\rho \mathrm{d} r.
    \label{eq:EigenfunctionFirstOrderCorrectionCoefficientsPolarContributions}
\end{equation}
This shows that the \emph{polar} first-order rotational corrections to the (polar) mode $\xi_\alpha^{(0)}$ are given by multipoles with the same $l$ and $m$.%
\footnote{Note that this applies only to the polar first-order rotational corrections. To obtain the \emph{axial} first-order rotational corrections to $\xi_\alpha^{(0)}$, one has to consider axial mode components in the mode sum \eqref{eq:EigenfunctionFirstOrderCorrectionExpansion}. In this case, as we will see below, the corrections are sourced by different multipoles.}

For later reference, we may also evaluate the eigenfrequency correction from \cref{eq:EigenfrequencyFirstOrderCorrection} as
\begin{equation}
    \omega_\alpha^{(1)} = 
        m_\alpha\Omega \,\mathcal{C}_\alpha \equiv
            \frac{m_\alpha\Omega}{\mathcal{A}_\alpha^{(0)2}}\int_0^R\left(2 W_\alpha^{(0)}V_\alpha^{(0)}+V_\alpha^{(0)2}\right)\rho \dd r,
    \label{eq:EigenfrequencyFirstOrderCorrectionPolarContributions}
\end{equation}
where $\mathcal{C}_\alpha$ is the \textit{Ledoux constant}.


In order to determine the axial components of the eigenfunction correction, we now take the inner product of the first-order eigenvalue equation \eqref{eq:QuadraticEigenvalueFirstOrder} with $\epsilon^{ijk}\nabla_j Y_{l'}^m\nabla_k r$. To evaluate certain spherical harmonic integrals, we make use of the recurrence relations
\begin{gather}
    \sin\theta\, \partial_\theta Y_l^m = l\mathcal{Q}_{l+1}^m Y_{l+1}^m - (l+1)\mathcal{Q}_l^m Y_{l-1}^m \\
    \intertext{and}
    \cos\theta\, Y_l^m = \mathcal{Q}_{l+1}^m Y_{l+1}^m + \mathcal{Q}_l^m Y_{l-1}^m, \\
    \intertext{where}
    \mathcal{Q}_l^m = \left[ \frac{(l-m)(l+m)}{(2l-1)(2l+1)}\right]^{1/2}
\end{gather}
(note that $\mathcal{Q}_l^m$ vanishes for $l=\pm m$). Then, we get
\begin{equation}
    \omega^{(0)2}_\alpha l'(l'+1) U^{(1)}_{\alpha l'} =
        2\Omega\omega^{(0)}_\alpha \left\{\delta_{l',l-1}\, l' \mathcal{Q}_{l'+1}^m \left[W^{(0)}_\alpha+(l'+2)V^{(0)}_\alpha\right]-\delta_{l',l+1}\, (l'+1) \mathcal{Q}_{l'}^m \left[W^{(0)}_\alpha-(l'-1)V^{(0)}_\alpha\right]\right\},
        \label{eq:QuadraticEigenvalueFirstOrderAxial}
\end{equation}
from which we obtain the axial contributions to the eigenfunction correction as
\begin{align}
    U^{(1)}_{\alpha,\,l-1} & = \frac{2\Omega}{\omega^{(0)}_\alpha}\frac{\mathcal{Q}_l^m}{l}\left[W^{(0)}_\alpha+(l+1)V^{(0)}_\alpha\right]
    \label{eq:QuadraticEigenvalueFirstOrderAxialMinusOnePiece} \\
    \intertext{and}
    U^{(1)}_{\alpha,\,l+1} & = -\frac{2\Omega}{\omega^{(0)}_\alpha}\frac{\mathcal{Q}_{l+1}^m}{l+1}\left(W^{(0)}_\alpha-l V^{(0)}_\alpha\right).
    \label{eq:QuadraticEigenvalueFirstOrderAxialPlusOnePiece}
\end{align}

\begin{figure}
    \centering
    \includegraphics[width=0.3\textwidth]{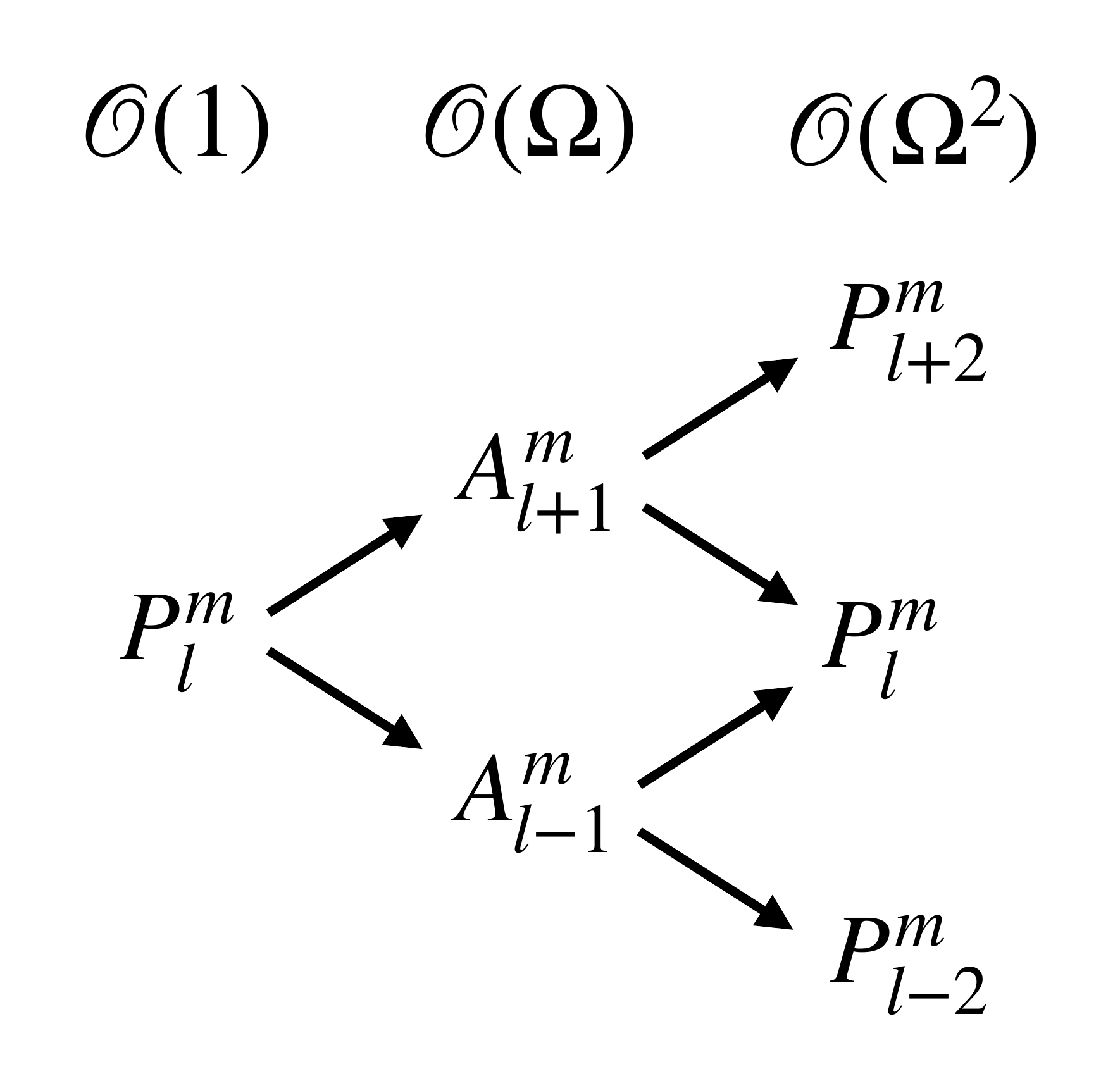}
    \caption{The hierarchy of multipole couplings in the slow-rotation calculation, assuming a polar mode in a non-rotating star.}
    \label{fig:modecoupling}
\end{figure}

Finally, the first-order correction to the polar mode eigenfunction is
\begin{equation}
    \xi_\alpha^{(1)i} 
        = \frac{W_{\alpha l}^{(1)}}{r}Y_l^m \nabla^i r
            + V_{\alpha l}^{(1)} \nabla^i Y_l^m
            - i U^{(1)}_{\alpha,\, l-1} \epsilon^{ijk}\nabla_j Y_{l-1}^m\nabla_k r
            - i U^{(1)}_{\alpha,\, l+1} \epsilon^{ijk}\nabla_j Y_{l+1}^m\nabla_k r.
    \label{eq:EigenfunctionFirstOrderCorrection}
\end{equation}
\Cref{eq:EigenfunctionFirstOrderCorrection} shows that, at first order in rotation, multipole couplings are introduced via axial pieces belonging to the neighbouring multipole orders \citep{1989nos..book.....U,2010aste.book.....A}. At the next order in rotation, these pieces couple back to their neighbouring polar multipole components, and so on (see \cref{fig:modecoupling}). The axial pieces $U^{(1)}_{\alpha,\,l-1}$ and $U^{(1)}_{\alpha,\,l+1}$ are given by \cref{eq:QuadraticEigenvalueFirstOrderAxialMinusOnePiece,eq:QuadraticEigenvalueFirstOrderAxialPlusOnePiece} respectively. The horizontal polar piece $V^{(1)}_{\alpha l}$ can be obtained from \cref{eq:QuadraticEigenvalueFirstOrderHorizontalPolar}, but one first needs to evaluate $(\delta p^{(1)}_{\alpha l}/\rho + \delta\Phi^{(1)}_{\alpha l})$. This, together with the radial piece $W^{(1)}_{\alpha l}$, can be calculated by formulating the first-order eigenvalue equation \eqref{eq:QuadraticEigenvalueFirstOrder} as a boundary value problem \citep{1981ApJ...244..299S}. From \cref{eq:QuadraticEigenvalueFirstOrderRadial2,eq:QuadraticEigenvalueFirstOrderHorizontalPolar}, and by also using the perturbed continuity equation \eqref{eq:PerturbedContinuity}, the perturbed Poisson's equation \eqref{eq:PerturbedPoissons}, and the equation of state for the perturbations \eqref{eq:PerturbedEoS2}, we obtain the system of equations
\begin{equation}
    \partial_r\left(\frac{W^{(1)}_{\alpha l}}{r}\right) =
        -\left(\frac{\partial_r p}{\Gamma_1 p}+\frac{2}{r}\right)\frac{W^{(1)}_{\alpha l}}{r}
        +\left[\frac{l(l+1)}{r^2\omega^{(0)2}_\alpha}-\frac{\rho}{\Gamma_1 p}\right]\left(\frac{\delta p^{(1)}_{\alpha l}}{\rho}+\delta\Phi^{(1)}_{\alpha l}\right)
        +\frac{\rho}{\Gamma_1 p}\delta\Phi^{(1)}_{\alpha l}
        +\frac{2}{r^2\omega^{(0)}_\alpha}\left[m\Omega\left(W^{(0)}_\alpha+V^{(0)}_\alpha\right)-l(l+1)\omega^{(1)}_\alpha V^{(0)}_\alpha\right],
    \label{eq:QuadraticEigenvalueFirstOrderRadialCorrection}
\end{equation}
\begin{equation}
    \partial_r\left(\frac{\delta p^{(1)}_{\alpha l}}{\rho}+\delta\Phi^{(1)}_{\alpha l}\right) =
        \left(\omega^{(0)2}_\alpha-\frac{A}{\rho}\partial_r p\right)\frac{W^{(1)}_{\alpha l}}{r}
        -A\left(\frac{\delta p^{(1)}_{\alpha l}}{\rho}+\delta\Phi^{(1)}_{\alpha l}\right)
        +A\delta\Phi^{(1)}_{\alpha l}
        +\frac{2\omega^{(0)}_\alpha}{r}\left(\omega^{(1)}_\alpha W^{(0)}_\alpha-m\Omega V^{(0)}_\alpha\right),
    \label{eq:QuadraticEigenvalueFirstOrderPressureAndPotentialCorrection}
\end{equation}
\begin{equation}
    \partial_r^2\delta\Phi^{(1)}_{\alpha l}+\frac{2}{r}\partial_r\delta\Phi^{(1)}_{\alpha l} =
        -4\pi G\rho A\frac{W^{(1)}_{\alpha l}}{r}
        +4\pi G\frac{\rho^2}{\Gamma_1 p}\left(\frac{\delta p^{(1)}_{\alpha l}}{\rho}+\delta\Phi^{(1)}_{\alpha l}\right)
        +\left[\frac{l(l+1)}{r^2}-4\pi G\frac{\rho^2}{\Gamma_1 p}\right]\delta\Phi^{(1)}_{\alpha l},
    \label{eq:QuadraticEigenvalueFirstOrderPotentialCorrection}
\end{equation}
which, accompanied by the appropriate boundary conditions, can be solved for the variables $W^{(1)}_{\alpha l}/r,\,\delta p^{(1)}_{\alpha l}/\rho + \delta\Phi^{(1)}_{\alpha l}$, and $\delta\Phi^{(1)}_{\alpha l}$.

As can be seen from \cref{eq:EigenfunctionFirstOrderCorrectionCoefficientsPolarContributions,eq:EigenfrequencyFirstOrderCorrectionPolarContributions}, as well as from \cref{eq:QuadraticEigenvalueFirstOrderRadialCorrection,eq:QuadraticEigenvalueFirstOrderPressureAndPotentialCorrection,eq:QuadraticEigenvalueFirstOrderPotentialCorrection}, the eigenfrequency and eigenfunction corrections $\omega^{(1)}_\alpha$ and $\xi^{(1)}_\alpha$ are directly proportional to $m\Omega$. Hence, the corrections can be simply obtained for some arbitrary values of $m$ and $\Omega$ and then rescaled for any other case.


\subsection{Mode normalisation} \label{subsec:Mode normalisation}

Based on the arguments from the previous two sections, we are finally in a position where we can make a more precise statement about the dynamical tide in a rotating star. In essence, we can make \cref{eq:RotatingLove} more tangible. As shown in \cref{sec:Normal modes of oscillation}, the modes of rotating stars obey a modified orthogonality relation, given by \cref{eq:RotatingOrthogonalityCondition}. Expanding all quantities to first order in $\Omega$ and separating the different orders, we obtain
\begin{gather}
	\left(\omega_\alpha^{(0)}+\omega_\beta^{(0)}\right)\left\langle\xi_\alpha^{(0)},\xi_\beta^{(0)}\right\rangle =
	\mathcal{B}_\alpha^{(0)}\delta_{\alpha\beta} \label{eq:RotatingOrthogonalityConditionZerothOrder} \\
    \intertext{and}
	\left(\omega_\alpha^{(0)}+\omega_\beta^{(0)}\right)\left[\left\langle\xi_\alpha^{(0)},\xi_\beta^{(1)}\right\rangle+\left\langle\xi_\alpha^{(1)},\xi_\beta^{(0)}\right\rangle\right]+\left(\omega_\alpha^{(1)}+\omega_\beta^{(1)}\right)\left\langle\xi_\alpha^{(0)},\xi_\beta^{(0)}\right\rangle-\left\langle\xi_\alpha^{(0)},i B^{(1)}\xi_\beta^{(0)}\right\rangle =
	\mathcal{B}_\alpha^{(1)}\delta_{\alpha\beta},
	\label{eq:RotatingOrthogonalityConditionFirstOrder}
\end{gather}
where the constants $\mathcal{B}_\alpha^{(0)}$ and $\mathcal{B}_\alpha^{(1)}$ will be fixed by the chosen mode normalisation. Using \cref{eq:NonRotatingOrthogonalityCondition}, the zeroth-order piece \eqref{eq:RotatingOrthogonalityConditionZerothOrder} gives
\begin{equation}
    \mathcal{B}_\alpha^{(0)} = 2\omega_\alpha^{(0)}\mathcal{A}_\alpha^{(0)2},
    \label{eq:RotatingOrthogonalityConditionZerothOrder2}
\end{equation}
namely, the zeroth-order constant $\mathcal{B}_\alpha^{(0)}$ is simply determined by the normalisation of the modes in the non-rotating case. A common choice is, for instance, $\mathcal{A}_\alpha^{(0)2}=MR^2$. From the first-order piece \eqref{eq:RotatingOrthogonalityConditionFirstOrder}, using \cref{eq:EigenfrequencyFirstOrderCorrection}, we get
\begin{equation}
    \mathcal{B}_\alpha^{(1)} = 2\omega_\alpha^{(0)}\left[\left\langle\xi_\alpha^{(0)},\xi_\alpha^{(1)}\right\rangle+\left\langle\xi_\alpha^{(1)},\xi_\alpha^{(0)}\right\rangle\right] \equiv
    2\omega_\alpha^{(0)}\mathcal{A}_\alpha^{(1)},
    \label{eq:RotatingOrthogonalityConditionFirstOrder2}
\end{equation}
with the constant $\mathcal{A}_\alpha^{(1)}$ depending on the normalisation of the first-order eigenfunction correction.

In order to understand how the normalisation of the rotationally-corrected eigenfunctions works, we should first note that the general solution to the first-order eigenvalue equation \eqref{eq:QuadraticEigenvalueFirstOrder} has the form $\xi^{(1)}_\alpha = q^{(1)}_\alpha \xi^{(0)}_\alpha + \tilde{\xi}^{(1)}_\alpha$, where $q^{(1)}_\alpha$ is an arbitrary constant. This is because, comparing to the zeroth-order eigenvalue equation \eqref{eq:NonRotatingQuadraticEigenvalue}, we see that the solution comprises a homogeneous and an inhomogeneous piece. The same can also be seen from the expansion of the eigenfunction correction in terms of the modes of the non-rotating star \eqref{eq:EigenfunctionFirstOrderCorrectionExpansion}: the component along $\xi_\alpha^{(0)}$ is given by the coefficient $c_{\alpha\alpha}^{(1)}$, which cannot be obtained from \cref{eq:EigenfunctionFirstOrderCorrectionCoefficients} and is thus left undetermined. Therefore, making a choice for the constant $\mathcal{A}_\alpha^{(1)}$ fixes the component of the first-order eigenfunction correction along $\xi_\alpha^{(0)}$, because $q^{(1)}_\alpha = c^{(1)}_{\alpha\alpha} =  \langle\xi^{(0)}_\alpha,\xi^{(1)}_\alpha\rangle/\mathcal{A}^{(0)2}_\alpha$. Using \cref{eq:EigenfunctionPolar,eq:EigenfunctionFirstOrderCorrection}, we get
\begin{equation}
    \mathcal{A}^{(1)}_\alpha = 
        2\int_0^R \left[ W_\alpha^{(0)}W_{\alpha l}^{(1)} + l(l+1) V_\alpha^{(0)}V_{\alpha l}^{(1)} \right]\rho\dd r.
\end{equation}

In this case, a common choice in perturbation theory (\textit{e.g.}, see \citealt{1970mmp..book.....M}) is to set $\mathcal{A}^{(1)}_\alpha=0$, \textit{i.e.}, to eliminate the component of $\xi^{(1)}_\alpha$ along $\xi^{(0)}_\alpha$ altogether. However, in order to demonstrate how the chosen normalisation enters the calculation and, ultimately, does not affect the final result, we will leave $\mathcal{A}^{(1)}_\alpha$ unspecified in the following.


\subsection{The Love number} \label{subsec:The Love number}

Based on the above, we can obtain the rotational corrections to the Love number by expanding \cref{eq:RotatingLove} in terms of $\Omega$ and hence working out---for the first time---an expression for the complete mode sum of a (slowly) rotating star. We have
\begin{equation}
    I_{\alpha' l} = I_{\alpha'}^{(0)} + I_{\alpha' l}^{(1)} + \mathcal{O}\left(\Omega^2\right)
\end{equation}
and
\begin{equation}
    \mathcal{B}_{\alpha'} = \mathcal{B}_{\alpha'}^{(0)} + \mathcal{B}_{\alpha'}^{(1)} + \mathcal{O}\left(\Omega^2\right),
\end{equation}
with the constants $\mathcal{B}_{\alpha'}^{(0)}$ and $\mathcal{B}_{\alpha'}^{(1)}$ given by \cref{eq:RotatingOrthogonalityConditionZerothOrder2,eq:RotatingOrthogonalityConditionFirstOrder2} respectively. Finally, the eigenfrequencies are given by 
\begin{equation}
    \omega_{\alpha'} 
        = \omega_{\alpha'}^{(0)} + m \, \mathcal{C}_{\alpha'} \Omega + \mathcal{O}\left(\Omega^2\right),
\end{equation}
with $\mathcal{C}_{\alpha'}$ defined in \cref{eq:EigenfrequencyFirstOrderCorrectionPolarContributions}. As previously discussed, the modes of a non-rotating, spherically symmetric star are degenerate in the order $m$. Rotation splits this degeneracy at $\mathcal{O}(\Omega)$. The mode that moves in the opposite direction to $\alpha'$ with the same $|m|$ has the rotating-frame frequency 
\begin{equation}
    \omega_{\beta'} 
        = \omega_{\alpha'}^{(0)} - m \, \mathcal{C}_{\alpha'} \Omega + \mathcal{O}\left(\Omega^2\right)
\end{equation}
(remember that we flip the sign of $m$ when we re-label these modes as $\beta'$; see \cref{subsec:Rotating stars}).

Based on the arguments above, at first order in rotation, only the term with $l'=l$ is needed in the sum over $l'$ in \cref{eq:RotatingLove}. Thus, we have
\newlength{\plussignlength}
\settowidth{\plussignlength}{+}
\begin{align}
    k_{lm}
        & = \frac{2 \pi G}{(2 l + 1) R^{2 l + 1}} \Bigg[
                \sum_{\alpha'} 
                    \frac{I_{\alpha'}^{(0)2}+2I_{\alpha'}^{(0)}I_{\alpha' l}^{(1)}}
                    {2\omega_{\alpha'}^{(0)}\mathcal{A}_{\alpha'}^{(0)2}}\left(1-\frac{\mathcal{A}_{\alpha'}^{(1)}}{\mathcal{A}_{\alpha'}^{(0)2}}\right)
                    \frac{1}{\omega_{\alpha'}^{(0)}+m\mathcal{C}_{\alpha'}\Omega+m\bar{\Omega}}
            \notag \\[5pt]
        & \phantom{=\frac{2 \pi G}{(2 l + 1) R^{2 l + 1}} \Bigg[}\hspace{-\plussignlength} +
                \sum_{\beta'} 
                    \frac{I_{\beta'}^{(0)2}+2I_{\beta'}^{(0)}I_{\beta' l}^{(1)}}
                    {2\omega_{\beta'}^{(0)}\mathcal{A}_{\beta'}^{(0)2}}\left(1-\frac{\mathcal{A}_{\beta'}^{(1)}}{\mathcal{A}_{\beta'}^{(0)2}}\right)
                    \frac{1}{\omega_{\beta'}^{(0)}-m\mathcal{C}_{\beta'}\Omega-m\bar{\Omega}}
        \Bigg] + \mathcal{O}\left(\Omega^2\right)
        \notag \\[5pt]
        & = \frac{2 \pi G}{(2 l + 1) R^{2 l + 1}} \left\{\rule{0cm}{.8cm}\right.
                \sum_{\alpha'} \left[
                    \frac{I_{\alpha'}^{(0)2}}{\mathcal{A}_{\alpha'}^{(0)2}\left[\omega_{\alpha'}^{(0)2}-\left(m\mathcal{C}_{\alpha'}\Omega+m\bar{\Omega}\right)^2\right]}
                    + \frac{I_{\alpha'}^{(0)2}}{\mathcal{A}_{\alpha'}^{(0)2}\omega_{\alpha'}^{(0)}\left[\omega_{\alpha'}^{(0)}+m\mathcal{C}_{\alpha'}\Omega+m\bar{\Omega}\right]}
                    \left(\frac{I_{\alpha' l}^{(1)}}{I_{\alpha'}^{(0)}}-\frac{\mathcal{A}_{\alpha'}^{(1)}}{2\mathcal{A}_{\alpha'}^{(0)2}}\right)
                \right]
            \notag \\[5pt]
        & \phantom{=\frac{2 \pi G}{(2 l + 1) R^{2 l + 1}} \Bigg\{}\hspace{-\plussignlength} +
                \sum_{\beta'}
                    \frac{I_{\beta'}^{(0)2}}{\mathcal{A}_{\beta'}^{(0)2}\omega_{\beta'}^{(0)}\left[\omega_{\beta'}^{(0)}-m\mathcal{C}_{\beta'}\Omega-m\bar{\Omega}\right]}
                    \left(\frac{I_{\beta' l}^{(1)}}{I_{\beta'}^{(0)}}-\frac{\mathcal{A}_{\beta'}^{(1)}}{2\mathcal{A}_{\beta'}^{(0)2}}\right)
        \left.\rule{0cm}{.8cm}\right\} + \mathcal{O}\left(\Omega^2\right).
        \label{eq:LoveNumberFirstOrderGeneralWithResonance}
\end{align}
In order to group the zeroth-order terms together, we used the same normalisation for the modes in the non-rotating limit, namely $\mathcal{A}^{(0)2}_{\beta'}=\mathcal{A}^{(0)2}_{\alpha'}$ and then took into account the fact that $I_{\beta'}^{(0)}=I_{\alpha'}^{(0)}$. Note that we have not (yet) expanded the denominators containing the mode frequencies, in order to keep the orbital resonances explicit.

\Cref{eq:LoveNumberFirstOrderGeneralWithResonance} is valid for any normalisation choice of the first-order corrections. Based on the discussion in \cref{subsec:Mode normalisation} though, we can expand the correction to the multipole moment into its homogeneous and inhomogeneous pieces as
\begin{equation}
    I_{\alpha' l}^{(1)} = \frac{\mathcal{A}_{\alpha'}^{(1)}}{2\mathcal{A}_{\alpha'}^{(0)2}} I_{\alpha'}^{(0)} + \tilde{I}_{\alpha' l}^{(1)}.
\end{equation}
This gives
\begin{align}
    k_{lm}
        = \frac{2 \pi G}{(2 l + 1) R^{2 l + 1}} \left\{\rule{0cm}{.8cm}\right.
                & \sum_{\alpha'} \left[
                    \frac{I_{\alpha'}^{(0)2}}{\mathcal{A}_{\alpha'}^{(0)2}\left[\omega_{\alpha'}^{(0)2}-\left(m\mathcal{C}_{\alpha'}\Omega+m\bar{\Omega}\right)^2\right]}
                    + \frac{I_{\alpha'}^{(0)}\tilde{I}_{\alpha' l}^{(1)}}{\mathcal{A}_{\alpha'}^{(0)2}\omega_{\alpha'}^{(0)}\left[\omega_{\alpha'}^{(0)}+m\mathcal{C}_{\alpha'}\Omega+m\bar{\Omega}\right]}
                \right]
            \notag \\[5pt]
            + & \sum_{\beta'}
                    \frac{I_{\beta'}^{(0)}\tilde{I}_{\beta' l}^{(1)}}{\mathcal{A}_{\beta'}^{(0)2}\omega_{\beta'}^{(0)}\left[\omega_{\beta'}^{(0)}-m\mathcal{C}_{\beta'}\Omega-m\bar{\Omega}\right]}
        \left.\rule{0cm}{.8cm}\right\} + \mathcal{O}\left(\Omega^2\right).
\end{align}
This shows that, in fact, only the inhomogeneous piece of the multipole moment contributes to the first-order correction. This result is equivalent to the one we would get had we eliminated the homogeneous piece from the start, by normalising the first-order solution such that $\mathcal{A}_{\alpha'}^{(1)}=0$.

The last step is to note that $\tilde{I}_{\beta' l}^{(1)}=-\tilde{I}_{\alpha' l}^{(1)}$, because, as discussed in \cref{subsec:The multipole expansion approach}, first-order corrections scale with $m\Omega$. Hence, we finally obtain
\begin{equation}
    k_{lm}
        = \frac{2 \pi G}{(2 l + 1) R^{2 l + 1}}
                \sum_{\alpha'}
                    \frac{I_{\alpha'}^{(0)2}}{\mathcal{A}_{\alpha'}^{(0)2}\left[\omega_{\alpha'}^{(0)2}-\left(m\mathcal{C}_{\alpha'}\Omega+m\bar{\Omega}\right)^2\right]}
                    \left(1-2m\Omega_\text{orb}\frac{\tilde{I}_{\alpha' l}^{(1)}}{\omega_{\alpha'}^{(0)}I_{\alpha'}^{(0)}}\right)
                + \mathcal{O}\left(\Omega^2\right).
        \label{eq:LoveNumberFirstOrderWithResonance}
\end{equation}
Alternatively, to get the formally correct slow-rotation expansion, we should also expand the denominator in terms of $\Omega$, which gives
\begin{equation}
    k_{lm}
        = \frac{2 \pi G}{(2 l + 1) R^{2 l + 1}}
                \sum_{\alpha'}
                    \frac{I_{\alpha'}^{(0)2}}{\mathcal{A}_{\alpha'}^{(0)2}\left[\omega_{\alpha'}^{(0)2}-\left(m\Omega_\text{orb}\right)^2\right]}
                    \left[1+2m\Omega_\text{orb}\left(\frac{(C_{\alpha'}-1)m\Omega}{\omega_{\alpha'}^{(0)2}-\left(m\Omega_\text{orb}\right)^2}-\frac{\tilde{I}_{\alpha' l}^{(1)}}{\omega_{\alpha'}^{(0)}I_{\alpha'}^{(0)}}\right)\right]
                + \mathcal{O}\left(\Omega^2\right).
        \label{eq:LoveNumberFirstOrder}
\end{equation}
From \cref{eq:LoveNumberFirstOrder} it is easy to see that, in the static limit $(\Omega_\text{orb}\to 0)$, we have
\begin{equation}
    k_l 
        = \frac{2 \pi G}{(2 l + 1) R^{2 l + 1}}
                \sum_{\alpha'}
                    \frac{I_{\alpha'}^{(0)2}}{\omega_{\alpha'}^{(0)2}\mathcal{A}_{\alpha'}^{(0)2}}
                + \mathcal{O}\left(\Omega^2\right).
        \label{eq:StaticLoveNumberFirstOrder}
\end{equation}

The final expressions for the effective Love number, \cref{eq:LoveNumberFirstOrderWithResonance,eq:LoveNumberFirstOrder}, show that the dynamical tidal response should be affected by the star's rotation already at first order, due to the corrections to the mode frequencies and the associated mass multipole moments. Given that, away from other orbital resonances, the $f$ mode is expected to make the largest contribution to the tidal response \citep{2020PhRvD.101h3001A}, one may choose to retain only the term corresponding to the appropriate $(l,m)$ $f$-mode multipole in the mode-sum representation of the Love number \citep{2021MNRAS.503..533A}. This case was considered in the study of the dynamical tides of Jupiter in \citet{2021PSJ.....2..122L}, and the  result can be reproduced from \cref{eq:LoveNumberFirstOrderWithResonance}, if only the $f$ mode is considered and one assumes that $\tilde{I}_f^{(1)}\ll I_f^{(0)}$ (which can be shown to be a good approximation for $f$ modes).%
\footnote{Note that in \citet{2021PSJ.....2..122L} and in \citet{2022ApJ...925..124D} the chosen mode normalisation implies that $\mathcal{B}^{(1)}_{\alpha'}=0$. However, this choice (or any other choice for that matter) does not imply, as mentioned in \citet{2022ApJ...925..124D}, that the first-order correction to the multipole moment $I^{(1)}_{\alpha' l}$ vanishes identically, because the inhomogeneous piece $\tilde{I}^{(1)}_{\alpha' l}$ is non-zero. What does vanish by making this normalisation choice is the combined first-order correction to the multipole moment of both the prograde and the retrograde mode, owing to the fact that $\tilde{I}_{\beta' l}^{(1)}=-\tilde{I}_{\alpha' l}^{(1)}$.}
For neutron star binaries, the modelling of resonantly-excited $f$ modes is also motivated by the fact that current gravitational-wave interferometers may be able to observe the late inspiral, where the $f$ mode orbital excitation  dominates the dynamical tide \citep{2016PhRvL.116r1101H, 2016PhRvD..94j4028S, 2019PhRvD.100b1501S, 2019PhRvD.100f3001V, 2020NatCo..11.2553P, 2022PhRvL.129h1102P, 2021PhRvR...3c3129S, 2022PhRvD.105l3032W, 2022PhRvD.106f4052K}. Our result for the effective Love number \eqref{eq:LoveNumberFirstOrderWithResonance} has the same functional form as the corresponding result in \citet{2021PhRvR...3c3129S}, which also includes a treatment of the near-resonance behaviour of the tidal response using the stationary-phase approximation, in order to be incorporated in waveform models. On the other hand, as seen from \cref{eq:StaticLoveNumberFirstOrder}, rotational corrections to the static Love number formally enter at second order in rotation.


\section{Discussion} \label{sec:Discussion}

The exciting era of gravitational-wave astronomy we find ourselves in promises to provide a vast number of future observations of binaries involving neutron stars. This presents an opportunity to constrain the equation of state of dense nuclear matter, currently an open question in nuclear physics and astrophysics. Tidal deformations in neutron star binaries induce phase shifts in the gravitational-wave signal, compared to that of two point particles, thus offering valuable information about their internal structure. This technique was already demonstrated with GW170817, the first direct detection of a neutron star binary with gravitational waves and the inaugural event for multimessenger astronomy.

So far, the majority of work on the subject has concentrated on the adiabatic treatment of the tidal deformations of compact binaries, which is relevant mainly at the early stages of the inspiral, where the binary evolution is slow. However, in the later stages, dynamical effects need to be accounted for. As the orbital evolution becomes faster, the tidal perturbation can no longer be treated as an instantaneous hydrostatic response. In addition, the star's normal oscillation modes are expected to be resonantly excited throughout the inspiral, as the orbital motion sweeps through the relevant frequencies. The high precision required by neutron star parameter estimation in order to adequately constrain the dense matter equation of state, in conjunction with the upcoming third-generation detectors where the sensitivity will be increased by an order of magnitude, motivate a comprehensive study of dynamical tidal effects in neutron star binaries.

In this paper, we have considered the dynamical tidal response of a spinning fluid star within the context of Newtonian gravity. The perturbation of a star due to an external gravitational field can be represented in terms of its normal modes. With the various classes of oscillation modes depending in different ways on the inner structure of the star, this description provides useful intuition about the impact that the different aspects of neutron star physics have on the problem. Rotation is one such aspect. Even though neutron star binaries which are close to merger are expected to have spun down and thus rotate slowly, rotation modifies the problem in several important ways, which are reflected on both the oscillation modes and the formalism itself: (i) it introduces a coupling among different multipoles and, as a result, a given mode can no longer be described by a single spherical harmonic $Y_l^m$; (ii) it lifts the degeneracy among modes with the same multipole degree $l$, but different azimuthal orders $m$, hence distinguishing prograde and retrograde modes; (iii) it gives rise to a new class of modes, the inertial modes, which need to be taken into account in the mode expansion of the tidal deformability; (iv) it deforms the background star into an oblate spheroid (at second order); (v) it modifies the mode orthogonality condition, necessitating the use of a different type of mode expansion (phase-space decomposition) in order to obtain uncoupled equations of motion for the modes.

We derived an expression for the effective tidal Love number using the mode-sum representation, both for the non-rotating and the rotating case, clarifying how the tidal response depends on the prograde and retrograde oscillation modes. We paid particular attention to the modifications to the formalism due to the presence of rotation, an issue which is often overlooked. In order to quantify the effect of rotation, we considered a slowly rotating body and carried out all the calculations at first order in the rotation, discussing the properties of mode corrections and addressing some subtle points related to mode normalisation. This led to an expression for the effective Love number which includes first-order corrections, due to changes in the mode frequencies and mass multipole moments. Finally, we demonstrated that the static tidal deformability is affected by rotation only at second order.

The influence of rotation on dynamical tides has been studied in the past, focusing on orbital mode resonances \citep{1997ApJ...490..847L, 1999MNRAS.308..153H, 2006PhRvD..74b4007L, 2017PhRvD..96h3005X, 2007PhRvD..75d4001F, 2017PhRvD..95d4023P, 2020PhRvD.101j4028P, 2020PhRvD.102f4059P, 2020PhRvD.101f4003B}, but also more recently in the context of gravitational waveform modelling, where the dynamical corrections induced by a resonantly-excited $f$ mode are considered \citep{2021PhRvR...3c3129S}. The importance of rotation in the description of dynamical tides was demonstrated lately, after the suggestion that the difference between the theoretically predicted and the observed value of Jupiter's tidal Love number may be explained by dynamical effects \citep{2021PSJ.....2...69I, 2021PSJ.....2..122L, 2022ApJ...925..124D, 2022PSJ.....3...11I, 2022PSJ.....3...89I}.

Moving forward, we need to bring these calculations into general relativity. Neutron stars are highly relativistic bodies and such calculations are required to incorporate realistic descriptions of the matter. There are, however, a number of issues that complicate this step. The problem becomes more complicated already in the post-Newtonian regime, as the gravito-magnetic tidal field couples to currents in the star (\textit{e.g.}, see \citealt{2020PhRvD.102f4059P, 2021PhRvD.103f4023P} for recent expositions of this aspect). 
Moreover, in general relativity, the modes are quasi-normal as the radiation of gravitational waves carries energy away from the star. For this reason, the modes have complex frequencies and no longer form a complete basis. Secondly, the back-scattering of waves by the curved background spacetime leads to the presence of late-time power-law tails. From the problem of perturbed black holes we know that the tail behaviour can be represented by a branch cut in the Green's function for the perturbation  \citep{1986PhRvD..34..384L, 1994PhRvD..49..883G, 1997PhRvD..55..468A}. The implications of this for the tidal problem have, as far as we are aware, not yet been considered. In addition, we need to make progress on the issue of how to represent a dynamical tidal field in general relativity. In short, there is a fair bit of work to be done.


\section*{Acknowledgements}

PP acknowledges support from the Mar\'ia Zambrano Fellowship Programme (ZAMBRANO21), funded by the Spanish Ministry of Universities and the University of Alicante through the European Union's ``Next Generation EU'' package, as well as from the grant PID2021-127495NB-I00, funded by MCIN/AEI/10.13039/501100011033 and by the European Union, from the Astrophysics and High Energy Physics programme of the Generalitat Valenciana ASFAE/2022/026, funded by the Spanish Ministry of Science and Innovation (MCIN) and the European Union's ``Next Generation EU'' package (PRTR-C17.I1), and from the Prometeo 2023 excellence programme grant CIPROM/2022/13, funded by the Ministry of Innovation, Universities, Science, and Digital Society of the Generalitat Valenciana. This work was also supported by the ``Ministero dell'istruzione, dell'universit\`a e della ricerca" (MIUR) PRIN 2017 programme (CUP: B88D19001440001), from the Amaldi Research Center, funded by the MIUR programme ``Dipartimento di Eccellenza" (CUP: B81I18001170001), and from the EU Horizon 2020 Research and Innovation Programme under the Marie Sk{\l}odowska-Curie Grant Agreement N. 101007855. FG, NA, and DIJ are  grateful for support from STFC via grant numbers ST/R00045X/1 and ST/V000551/1.


\section*{Data Availability}

Additional data related to this article will be shared on reasonable request to the corresponding author.


\bibliographystyle{mnras}
\bibliography{bibliography}

\bsp 
\label{lastpage}
\end{document}